\documentclass[apjl,twocolumn]{emulateapj_mod}
\usepackage{epsfig,apjfonts,mathptmx}

\def\gtsima{$\; \buildrel > \over \sim \;$}
\def\ltsima{$\; \buildrel < \over \sim \;$}
\def\prosima{$\; \buildrel \propto \over \sim \;$}
\def\gsim{\lower.5ex\hbox{\gtsima}}
\def\lsim{\lower.5ex\hbox{\ltsima}}
\def\simgt{\lower.5ex\hbox{\gtsima}}
\def\simlt{\lower.5ex\hbox{\ltsima}}
\def\simpr{\lower.5ex\hbox{\prosima}}

\def\h1{$h^{-1}$}
\def\ergpsec{erg~s${}^{-1}$}
\def\eeq{\end{equation}}
\def\beq{\begin{equation}}
\def\24mu{24\,$\mu$m}
\def\70mu{70\,$\mu$m}
\def\8mu{8\,$\mu$m}

\shorttitle{Compton thick AGNs inside Massive Star Forming Galaxies at \lowercase{z}~$\sim2$}
\shortauthors{E. Daddi et al.}

\begin{document}

\title{Multiwavelength study of massive galaxies at \lowercase{z}~$\sim2$.\\ 
II.  Widespread Compton thick AGN and the concurrent growth
of black holes and bulges}

\author{E. Daddi\altaffilmark{1}, 
D.M. Alexander\altaffilmark{2},
M. Dickinson\altaffilmark{3},
R. Gilli\altaffilmark{4},
A. Renzini\altaffilmark{5}, 
D. Elbaz\altaffilmark{1},
A. Cimatti\altaffilmark{6},
R. Chary\altaffilmark{7},
D. Frayer\altaffilmark{7},
F.E. Bauer\altaffilmark{8},
W.N. Brandt\altaffilmark{9}, 
M. Giavalisco\altaffilmark{10},
N.A. Grogin\altaffilmark{11},
M. Huynh\altaffilmark{7},
J. Kurk\altaffilmark{12},
M. Mignoli\altaffilmark{4}, 
G. Morrison\altaffilmark{13,14},
A. Pope\altaffilmark{15},
S. Ravindranath\altaffilmark{16} 
}

\altaffiltext{1}{Laboratoire AIM, CEA/DSM - CNRS - Universit\`e Paris Diderot,
DAPNIA/SAp, Orme des Merisiers,  91191 Gif-sur-Yvette, France {\em edaddi@cea.fr}}
\altaffiltext{2}{Department of Physics, Durham University, South Road,  Durham, DH1 3LE, UK}
\altaffiltext{3}{National Optical Astronomy Observatory,  950 N. Cherry Ave., Tucson, AZ, 85719, USA}
\altaffiltext{4}{INAF, Osservatorio Astronomico di Bologna, via Ranzani, 1 - 40127 Bologna, Italy}
\altaffiltext{5}{INAF, Osservatorio Astronomico di Padova, Vicolo Osservatorio 5, I-35122, Padova, Italy}
\altaffiltext{6}{Dipartimento di Astronomia, Universit\'a di Bologna, Via Ranzani 1, I-40127, Bologna, Italy}
\altaffiltext{7}{{\em Spitzer} Science Center, Caltech, MS 220-6, CA 91125, USA}
\altaffiltext{8}{Columbia Astrophysics Laboratory, Columbia University, Pupin Laboratories, 550 West 120th Street, Roomm 1418, New York, NY 10027}
\altaffiltext{9}{Department of Astronomy and Astrophysics, The 
Pennsylvania State University, 525 Davey Lab, University Park, PA 
16802, USA}
\altaffiltext{10}{University of Massachussets, Astronomy Department, Amherst, MA 01003, USA}
\altaffiltext{11}{School of Earth and Space Exploration, Arizona State University, Tempe, AZ 85287, USA}
\altaffiltext{12}{Max-Planck-Institut f\"ur Astronomie, K\"onigstuhl 17, D-69117, Heidelberg, Germany}
\altaffiltext{13}{Institute for Astronomy, University of Hawaii, Honolulu,
HI, 96822, USA}
\altaffiltext{14}{Canada-France-Hawaii Telescope, Kamuela, HI, 96743, USA}
\altaffiltext{15}{Department of Physics \& Astronomy, University of British Columbia, Vancouver, BC, V6T 1Z1, Canada}
\altaffiltext{16}{Inter-University Centre for Astronomy and Astrophysics, Pune University Campus, Pune 411007, Maharashtra, India}

\begin{abstract} 
Approximately 20--30\% of $1.4\simlt z\simlt 2.5$ galaxies with  $K_{\rm Vega}<22$ 
detected with Spitzer
MIPS at \24mu show excess mid-IR emission relative to that expected 
based on the rates of star formation measured from other multiwavelength data.  
These galaxies also display some near-IR excess in  Spitzer IRAC data, 
with a spectral energy distribution peaking longward of 1.6$\mu$m in the rest frame, 
indicating the presence of warm-dust emission usually absent in star forming galaxies.
Stacking {\it Chandra} data for the mid-IR excess galaxies yields a 
significant hard X-ray detection at rest-frame energies $>$~6.2~keV. The stacked 
X-ray spectrum rises steeply at $>10$~keV, suggesting that these sources host 
Compton-thick Active Galactic Nuclei (AGNs) with column densities 
$N_{\rm H}\simgt10^{24}$~cm${}^{-2}$ and an average, unobscured X-ray luminosity 
$L_{\rm 2-8 keV}\approx (1$--$4)\times 10^{43}$~\ergpsec.  Their sky density 
($\sim3200$~deg${}^{-2}$) and space density ($\sim2.6\times10^{-4}$~Mpc${}^{-3}$) 
are twice those of X-ray detected AGNs at $z\approx$~2, and much larger than those 
of previously-known Compton thick sources at similar redshifts.  The mid-IR excess 
galaxies are part of the long sought-after population of distant heavily obscured AGNs 
predicted by synthesis models of the X-ray background.  The fraction of mid-IR excess 
objects increases with galaxy mass, reaching $\sim50$--60\% for $M\sim10^{11}M_\odot$, 
an effect likely connected with downsizing in galaxy formation.  The ratio of the 
inferred black hole growth rate from these Compton-thick sources to the global star 
formation rate at $z=2$ is similar to the mass ratio of black holes to stars in 
local spheroids, implying concurrent growth of both within the precursors of 
today's massive galaxies.
\end{abstract}

\section{Introduction}

Active Galactic Nuclei (AGN) and galaxies are intimately connected. 
It is generally thought that virtually all nearby early type galaxies 
(and perhaps all relatively massive
galaxies) contain a supermassive central black hole (BH), indicating that all 
massive galaxies may have experienced an AGN phase in their past (Kormendy \& Richstone 1995; Magorrian et al.~1998). 
Moreover, remarkably tight correlations have now been unveiled
between central supermassive BH and the mass or velocity dispersion of 
the bulge in their host galaxies
(Ferrarese \& Merritt 2000; 
Gebhardt et al.~2000), with typical mass ratio 
$M_{\rm BH}/M_{\rm bulge}\sim(1$--$2)\times10^{-3}$ 
(e.g., McLure \& Dunlop 2001; Ferrarese et al 2006).  This suggests but does not establish that AGN and star 
formation activity may have been concurrent.
The origin of these correlations is in fact not entirely clear, 
and has been the subject of active debate during recent years.
It appears to be connected to the positive feedback that the 
BH can exercise on its host galaxy, 
becoming effective 
if the BH exceeds some critical mass limit and thus luminosity 
(Silk \& Rees 1998; di Matteo, Springel \& Hernquist 2005; 
Springel, di Matteo \& Hernquist 2005; Schawinski et al.~2006). 
If the energetic
emission from accretion around the BH is large enough, 
the galaxy can be cleared out of its gas
either via heating (e.g., 
Ciotti \& Ostriker 1997; 2007) or accretion related outflows as
in BAL quasars (e.g., Chartas et al. 2007), 
thereby preventing further gas accretion.

In order to probe BH growth in the distant universe
efficient tracers of AGN activity are needed.  Obtaining a complete census of 
active galaxies is complicated due to  
AGN obscuration by circumnuclear material. 
During the formation epochs of galaxies, it can be well expected that a
greater abundance
of gas and its chaotic motions due to merging 
and interactions cause the typical 
AGN obscuration to be much larger than locally.  In that case, a 
substantial fraction of the cosmic BH growth history 
could be hidden from view and not yet taken into account.
It is also currently unknown if the "Unification Paradigm" describing the AGN
phenomenon (e.g., Antonucci 1993;  Urry \& Padovani 1995) holds
at high redshift, in particular if the ratio of obscured 
to unobscured AGNs (or, more accurately, the distribution of
obscuring column densities) is the same as observed locally. 
More generally, the shape of the cosmic X-ray background, with its peak toward 30~keV (Marshall et al.~1980), suggests that
much of the BH-accretion luminosity at substantial redshifts is obscured (e.g.,
Fabian \& Iwasawa 1999). 

The deepest X-ray surveys available to date, performed with the {\it Chandra} satellite
in the {\it Chandra} Deep Field South (Giacconi et al.~2002; GOODS-South field, GOODS-S hereafter),  {\it Chandra} Deep Field North (Alexander
et al.~2003a; GOODS-North, GOODS-N hereafter), and with {\it XMM-Newton} 
(e.g., Hasinger~2004\ in the Lockman hole) resolved most of the 
X-ray background at energies
0.5--6~keV, but a substantial fraction of the higher energy background 
remains unresolved (Krivonos et al.~2005; Worsley et al.~2005; 2006; 
Hickox \& Markevitch 2007). Synthesis models have been built (Comastri et al.~1995; Gilli et al.~2001; 2007) showing that in order to account 
for the high energy X-ray background, where most of the accretion energy in the universe is being emitted, one has to postulate
the existence of substantial populations of high redshift heavily obscured
Compton thick AGNs (with column densities of 
$N_{\rm H}\simgt10^{24}$~cm${}^{-2}$). 
Such a population of Compton thick
sources at high redshift
still awaits discovery (e.g., Barger et al.~2007).
Hard X-ray selection, the best means currently to find such sources at 
high redshift, is limited by telescope sensitivities
and would allow one to find only the brightest sources and 
with non-extreme column 
densities (e.g., Brandt \& Hasinger 2005). As an example, the archetype 
Compton thick AGN NGC~1068, with $L_{\rm X}>10^{44}$~\ergpsec,
would have a hard X-ray (2--8~keV) flux of only 
$\sim7\times10^{-18}$~\ergpsec~cm${}^{-2}$ at $z=2$, more than
an order of magnitude fainter than that of sources detectable in the deepest 
(2~Ms) {\it Chandra} survey, in the GOODS-North field.

On the other hand, some of the energy absorbed by the obscuring
material is reprocessed and re-emitted at longer wavelengths,
providing the opportunity to detect these sources via their IR
emission. For example, the nucleus of NGC~1068 would have a flux density of 25$\mu$Jy at
\24mu    if placed at $z=2$ --- very faint but still detectable in the
ultradeep {\it Spitzer}~MIPS imaging taken as part of GOODS\@.  One of
the primary goals of GOODS was indeed the discovery of previously
unknown populations of distant AGNs (e.g., Treister et al.~2004) by
combining the deepest multiwavelength imaging brought to bear upon
these fields. Several recent studies have focused on finding distant
obscured AGNs by means of mid-IR or far-IR selection, in GOODS and
other deep {\it Spitzer} surveys.  For example, selection by means of
red IRAC (3.5--8$\mu$m) colors were presented by Lacy et al.~(2004),
Stern et al.~(2005), and similarly the selection by means of power-law
spectral energy distributions (SEDs) over IRAC bands (Alonso Herrero
et al.~2006; Donley et al.~2007) have provided AGN candidate samples
with only partial overlap with X-ray selected sources.

The peak of QSO activity has been established to take place at $z\approx2$
(e.g., Schmidt 1968; Croom et al. 2004; Hasinger et al. 2005). 
Therefore, it is quite natural to look for the growth
of BH in massive galaxies at the same epoch. Indeed, the massive 
$z\sim2$ galaxies selected at
submillimeter wavelengths (SMGs) are undergoing star-formation activity and at
the same time host luminous AGNs, just the signatures expected for joint
stellar-BH growth (e.g.,\ Alexander et~al. 2005a,b; Borys et~al.
2005). However, these sources are relatively rare and
appear to represent the most strongly
star-forming galaxies at these epochs (e.g.,\ Chapman et~al. 2005) and it
is important to see if this joint growth is also occurring in more typical
massive galaxies at $z\sim2$.

This paper describes a source population serendipitously discovered 
while studying the multiwavelength 
emission properties of distant $1.4<z<2.5$ massive star forming galaxies 
in the GOODS fields,
described in a companion paper (Daddi et al.~2007, \hbox{Paper~I} hereafter). 
As discussed in \hbox{Paper~I},
we have built $z\sim2$ galaxy samples free of detectable AGN activity. 
A sizable fraction shows distinct mid-IR excess 
with respect to that expected from star formation activity alone.
This property of $z\sim2$ galaxies implies that the extremely 
efficient and penetrating \24mu    photometry
from {\it Spitzer} is not always
a reliable estimator of the ongoing star formation rates (SFRs) in galaxies, 
especially in the case of sources with very bright mid-IR luminosities
(particularly those encountered at $z\sim2$ in surveys shallower than GOODS).
In this paper, we investigate the nature of these
mid-IR excess sources in detail. 
Stacking in the X-ray bands with {\it Chandra} clearly unveils heavily obscured
AGNs, in many cases (or perhaps most cases) with Compton thick central BHs,
which might be responsible for the mid-IR excess.
The \24mu    photometry is therefore still extremely valuable also in the
case of bright \24mu    galaxies at $z=2$, as it allows us
to investigate this previously unidentified population of distant AGNs,
relevant for the history of accretion onto super-massive BHs 
and their feedback on galaxy formation and evolution at a crucial 
epoch of massive galaxy formation.

The paper is organized as follows: in Section~\ref{sec:data} we
describe the mid-IR excess galaxy sample and summarize all relevant
background work described in full detail in the companion paper
(\hbox{Paper~I}).  Section ~\ref{sec:multi} compares the mid-IR to UV
estimated SFRs (i.e., the mid-IR excess) to a number of other observed
galaxy properties and shows that a warm dust component in addition to
the colder star formation component is present in these galaxies at
near-IR to mid-IR rest-frame wavelengths.  The X-ray stacking
analysis, unveiling the presence of heavily obscured AGNs, are
presented in Section~\ref{sec:X}, with the constraints on the
absorbing gas column density suggesting that these sources are
typically Compton thick. We discuss the implications of these results
with regard to the obscured AGN fraction at high redshifts, the X-ray
background, the coeval growth of BH and galaxies, and AGN duty cycle
in Section~\ref{sec:impli}; the possible relevance of these results for
AGN feedback is discussed in Section~\ref{sec:feed}. Future prospects for
further developments in understanding this previously unknown
population of $z\sim2$ AGNs are given in
Section~\ref{sec:future}. Summary and conclusions are provided in
Section~\ref{sec:end}.  We quote stellar masses and star formation
rates for the case of a Salpeter IMF from 0.1 and 100 $M_\odot$, and
we adopt a WMAP cosmology with $\Omega_\Lambda, \Omega_M = 0.74,
0.26$, and $h = H_0$[km s${}^{-1}$ Mpc${}^{-1}$]$/100=0.73$ (Spergel
et al.~2003).  Unless explicitly stated otherwise, we quote magnitudes
in the Vega scale.

\setcounter{footnote}{0}

\section{The Sample}
\label{sec:data}

\subsection{Galaxy selection and datasets}

The dataset and galaxy samples used in this paper are fully described in \hbox{Paper~I}.
For completeness we provide the relevant information here.
We have selected sources in the $K$-band to completeness limits of $K=20.5$ 
in the GOODS-North
and $K=22$ in the GOODS-South field, and then used the $BzK$ color 
selection technique of Daddi et al.~(2004b) to identify
a total of $\sim1200$ galaxies at $1.4\simlt z \simlt 2.5$. 
Accurate photometric redshifts 
($\sigma_z\sim0.25$)
as well as a large number of spectroscopic redshifts have been used. We exclude
from the analysis
the small fraction of contaminant $BzK$ galaxies with $z<1.2$ or $z>3$. 
Sources of spectroscopic redshifts 
are described in \hbox{Paper~I} and include all available public
datasets, 
plus currently unpublished redshifts from the ultradeep (15-30 hours per mask)
spectroscopy of the GMASS survey (Kurk et al.~2007, in preparation).
We have excluded AGN-hosting galaxies by eliminating from the sample (1) 
sources with a 2--8~keV band detection in the 1~Ms and 2~Ms {\it Chandra} Deep Field X-ray catalogs 
of Alexander et al.~(2003a) for GOODS-S and N respectively, and (2)
sources with power-law SEDs over {\it Spitzer}~IRAC and MIPS, 
lacking a distinct decrease in the flux 
density ($f_\nu$) at wavelengths beyond $1.6\mu$m rest-frame, based on visual 
classification (eventually, this removes only 6 galaxies in total
due to power-law SEDs that were not also detected in the hard X-ray band). 
We also excluded 10\% of the galaxies in the sample because of source blending due to the
low resolution ($\sim1.6$\arcsec) at IRAC
wavelengths, which generally hampers measurement of accurate
{\it Spitzer}~MIPS \24mu    flux densities as well.

The main aim of \hbox{Paper~I}  is to investigate the nature 
of star formation at high redshift and to identify the best mean to estimate
the ongoing SFR activity.  The work thus mostly focuses on $BzK$ {\em star forming} galaxies (as opposed to $BzK$ {\em passive} galaxies).
All the multiwavelength star formation rate (SFR) indicators available in GOODS
have been used.  These include: (1) the deepest current {\it Spitzer}~MIPS 
\24mu    imaging (Chary et al.~2007, in preparation), 
reaching about 12$\mu$Jy (3$\sigma$) in both fields; (2) the deepest current {\it Spitzer}~MIPS 70\,$\mu$m imaging reaching about 1.8mJy
in part of both GOODS fields (Frayer et al.~2006 and in preparation); 
(3) 160 hours of VLA 1.4~GHz data in GOODS-N with 4.7$\mu$Jy of rms per beam
(Morrison et al.~2007, in preparation); (4) 0.5--2~keV
X-ray imaging from {\it Chandra} (used mainly in this paper, where we also discuss the
AGN contamination that is needed for the interpretation of those data); (5)
a deep 850\,$\mu$m map of GOODS-N (Pope et al.~2005);
and (6) UV photometry from GOODS ground-based and HST imaging
(Giavalisco et al. 2004). 
The way in which photometry at these
various wavelengths is interpreted to measure SFR in galaxies, and the implied 
known limitations, are described in Section 3 of \hbox{Paper~I}. 
The main results of \hbox{Paper~I}, relevant to this work, 
are presented in Section 4 and 5 of that paper. 
In summary, we find that a fraction of the $z=2$
galaxies (generally those with the brightest mid-IR
luminosities)
have significant excess flux density in the mid-IR at \24mu,     
i.e.~at rest-frame $\sim8\mu$m, compared to the flux density expected on the
basis of their ongoing star formation activity. 
A consistent mid-IR excess is also
inferred from a comparison 
with radio, 70\,$\mu$m, UV, submm (\hbox{Paper~I}) (and soft X-ray,
as discussed later in this paper) derived SFRs. We also find that 
the SFRs estimated from the dust reddening-corrected UV luminosities
compares fairly well with respect to all other available indicators. 

The aim of this paper is therefore to carefully identify and study the
population of sources with mid-IR excess.
In order to work with meaningfully defined galaxy samples, being able to
compute accurately SFRs from both mid-IR and from the UV, 
we limited in the following the analysis to
\24mu    detected sources with well defined UV slopes
(i.e. those with  an error in $(B-z)$ color less than 0.4~mag). 
This is required in order to obtain reasonable estimates of dust corrected
UV SFR ($SFR_{\rm UV, corr}$), with a formal error in the  reddening correction
lower than a factor of $\approx 2$.

We further excluded galaxies that appear to be
quiescent, or have only a low level of star-formation activity.  These
sources are generally very red in the UV because of
relatively old stellar populations affecting the colors, and not simply
because of dust reddening.  For such sources the dust 
extinction correction is expected to fail, 
leading to overestimated $SFR_{\rm UV, corr}$, 
so that the comparison of mid-IR based SFRs to UV ones is not feasible.
Based on the \hbox{Paper~I} finding of a tight correlation between SFR and
stellar mass at redshift $z\sim2$ (as well known now to exist from 
$z=0$ to $z=1$, 
Elbaz et al.~2007; Noeske et al.~2007), we excluded quiescent galaxies
as those for which the
specific star formation rate $SSFR_{mid-IR}< SSFR_{mid-IR}^0/3$, 
where $SSFR \equiv SFR_{mid-IR}/Mass$ 
and $SSFR_{mid-IR}^0$ is the median for \24mu    detected $z\sim2$ 
galaxies in GOODS. This criterion also requires a \24mu    detection.

\subsection{Definition of mid-IR excess sources}

Following the results from \hbox{Paper~I}, we define here
mid-IR excess sources as sources satisfying

\beq
{\rm Log} (SFR_{mid-IR+UV}/SFR_{\rm UV, corr}) > 0.5 
\eeq

\noindent
Here $SFR_{mid-IR+UV}$ is derived from the \24mu    flux density at the measured
spectroscopic or photometric redshift, 
and using the Chary \& Elbaz (2001) templates, which are based
on local 
mid- to far-IR correlations, with the addition of $SFR$ estimated from the directly observed UV luminosity (non-reprocessed
light from star formation).  The term $SFR_{\rm UV, corr}$ is estimated from the UV after correcting for dust extinction based 
on the Calzetti et al.~(2000) law
(the average correction  is a factor of $\sim40$ for $K<20$ sources, Daddi 
et al.~2005b, and decrease to a factor of a few to the faintest $K=22$
limits).
A galaxy can satisfy this mid-IR excess 
criterion either from
overestimating the $SFR_{mid-IR}$ respect to $SFR_{UV}$, or from underestimating the
latter relative to the former. The first case corresponds to a genuine excess
flux density in the mid-IR; the second one can be due to dust obscuration exceeding
the Calzetti law prediction (optically-thick UV emission).
In \hbox{Paper~I} it is shown that the second case is relevant at most 
for a small fraction of the sample ($\simlt15$\%), 
and therefore the vast majority (but not
all) of mid-IR excess sources are overluminous, for some reason, at mid-IR
wavelengths. 

It is important to recall here that the luminosity
dependent Chary \& Elbaz (2001) templates that we adopt have 
been built on the basis of the local correlations in the mid-IR to far-IR
for star forming galaxies.
A mid-IR excess with respect to these SEDs is equivalent to a mid-IR excess 
with respect to the SED of standard star forming galaxies in the local 
universe. In any case, it should be clarified that this population of mid-IR 
excess galaxies is well defined independently of the class of SEDs used
to interpret the mid-IR luminosities. 
As shown extensively in \hbox{Paper~I}, $z=2$ galaxies
at a given intrinsic $SFR$ or bolometric luminosity appear to display
a wide range in their $L(8\mu$m). Therefore mid-IR excess galaxies can be 
also seen as galaxies with  $L(8\mu$m) larger than the average 
$<L(8\mu{\rm m})>$
of galaxies of a given SFR (or of local star forming 
galaxies with the same SFR). Deviations from the normal $L(8\mu$m) versus
SFR relation
are also observed in the local universe, and generally ascribed to the
presence of an AGN (e.g., Genzel et al. 1998; Laurent et al. 2000;
Dale et al. 2006; Armus et al.~2007).

\begin{figure}    
\centering 
\includegraphics[width=8.8cm]{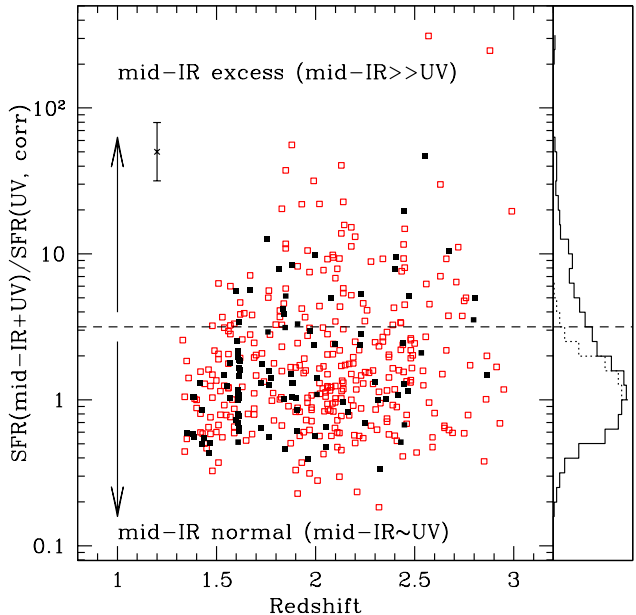} 
\caption{
The ratio of mid-IR-based to UV-based star formation rate (SFR)  
is plotted as a function of the redshift. Here and in several of 
the figures in this paper, we plot only 
the GOODS-S portion of the galaxy sample, being deeper and extending 
to $K<22$. Filled squares indicate spectroscopic
redshifts, empty squares indicate photometric redshifts. The horizontal line defines the
separation between mid-IR-excess and normal galaxies, as given by Eq.~1.
The rightmost panel shows the distribution of sources as a function
of mid-IR excess ratio. The dotted line is the reflection of the bottom 
part of the histogram around a ratio of 1. The error bar in the top-left 
part of the figure shows the typical error in the SFR ratio, inferred 
from the spread of the histogram around a ratio of 1.
}
\label{fig:zz_S}
\end{figure}

The threshold chosen in Eq.~1 (0.5~dex, about a factor of 3) is
dictated by the evidence that, in the opposite direction to mid-IR
excess galaxies, there are very few galaxies with $SFR_{\rm UV, corr}$ in
excess of more than a factor of 3 of $SFR_{mid-IR+UV}$
(Fig.\ref{fig:zz_S}).  This suggests that samples of genuinely
discrepant galaxies with too high ratios are to be found above a
similar threshold, assuming that intrinsic scatter in the ratio is to
first order symmetrical.\footnote{It is worthwhile to emphasize that,
despite the large dust reddening corrections to $SFR_{\rm UV, corr}$, most
sources in Fig.~\ref{fig:zz_S} cluster around a ratio of 1, and the
Gaussian component in the distribution has a dispersion of only
$\sim0.2$~dex} We will later show that this threshold is well defined
also with respect to the X-ray properties of the samples.

Although the use of UV as the primary SFR indicator is less than 
optimal because of the possibility for substantial obscuration from 
thick lines of sight,  it is the only one
available for nearly all \24mu    sources at $z\sim2$ in GOODS
(the other SFR indicators being limited by sensitivity and applicable 
individually to only 
a fraction of the galaxies, e.g., 20\% for radio). 

\begin{figure}    
\centering 
\includegraphics[width=8.8cm]{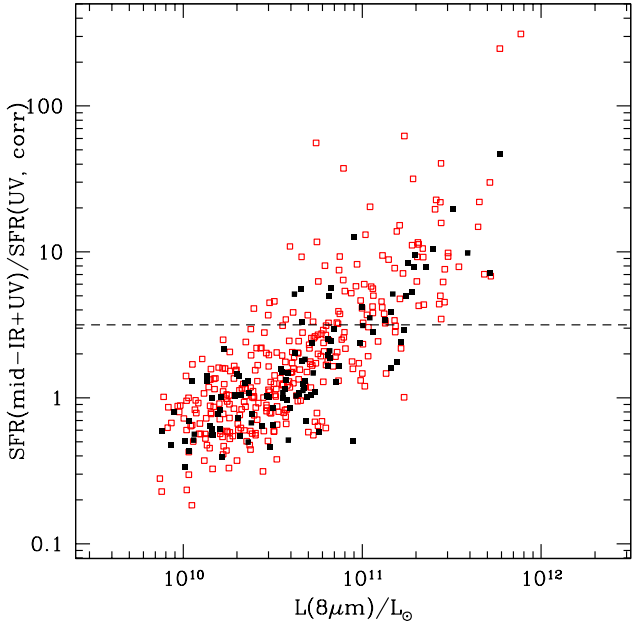} 
\caption{
The mid-IR excess is observed to correlate with the
mid-IR luminosity of galaxies (\hbox{Paper~I}; the lack of sources 
in the upper-left part of the plot might be in part a selection effect). 
Symbols are like in Fig.~\ref{fig:zz_S}.  The majority of the sources 
with $L(8\mu$m)$>10^{11}L_\odot$ would be classified as mid-IR excess
sources based on the definition in Eq.~1. 
Multiwavelength analysis in \hbox{Paper~I} (and later in this paper)
shows that the excess is present with respect to radio, 
UV, 70\,$\mu$m, soft X-ray, and submm SFR indicators.
}
\label{fig:L8}
\end{figure}

The combination of the requirements discussed in the previous section
leaves us with a total sample of 430 (183)
galaxies in the GOODS-S (GOODS-N) field, or about 50\% of the primary sample. 
Of these, 106 (58) satisfy Eq.~1 and are therefore classified as mid-IR excess
galaxies --- representing 24.6\% (31.7\%) of the total sample. 
Likewise, robust spectroscopic redshifts
are available for 108 (44) galaxies, or 25.1\%
(24.0\%) of the total sample.
Most of the mid-IR excess sources have very high $8\mu$m rest-frame 
luminosities, often in excess of $10^{11}L_\odot$ (Fig.\ref{fig:L8}), and the 
vast majority of galaxies in our sample having $L(8\mu$m)$>10^{11}L_\odot$ 
also shows
a mid-IR excess. This is reminiscent of finding that most
luminous IR galaxies in the local universe tend to be AGN dominated (e.g., Tran et al.~2001).

\section{Mid-IR excess and other galaxy properties}
\label{sec:multi}

With the aim of understanding the nature of the sources with mid-IR excess, 
and therefore the origin of this excess,
we have searched for the existence of possible correlations of this excess 
with other galaxy properties. 

\subsection{UV slopes, redshift, morphology and colors}

As shown in Fig.~13 of \hbox{Paper~I}, the mid-IR excess does not 
strongly depend on the UV slope of the galaxies, 
and we therefore exclude that it is due to effects inherent 
in the correction for reddening of the UV luminosity.
A possible origin of the excess could instead be mid-IR SED evolution with 
redshift, within the $1.2<z<3$ range populated by our sample.
We have checked if mid-IR excess sources preferentially appear at some 
high or low redshift within the $z\sim2$ sample (Fig.~\ref{fig:zz_S}).
There is a tendency for $z\sim1.5$ sources to have low mid-IR to UV ratios. 
This could be due to 
9.7$\mu$m silicate absorption entering the \24mu    
{\it Spitzer}~MIPS bandpass,
or to an overestimation of the reddening correction at these lowest redshifts. 
There is also a tendency for sources with the strongest excess to be located at 
$z\simgt2$. However, the median ratio of mid-IR to UV based SFRs is fairly 
stable  as a function of redshift, and no strong trend with redshift is 
detected. 
If we are seeing galaxies with anomalous mid-IR SEDs, compared
to $z\sim0$, they are rather homogeneously distributed in the probed redshift
range.

We have performed Sersic index fitting using Galfit (Peng et al. 2002) to the 
ACS F850W band images of our
sample of $z=2$ galaxies in GOODS-S. 
Results and full details will be presented in 
Ravindranath et al. (in preparation). Comparing the Sersic index and sizes
distributions of mid-IR excess and normal galaxies, we do not find statistically
significant differences for the full sample with $K<22$ in GOODS-S.

Instead, we find a strong correlation of the mid-IR excess with the $(K-IRAC)$ colors. Fig.~\ref{fig:SFR_ch4}
shows that mid-IR excess objects have redder ($K-5.8\mu$m) colors. Galaxies
with spectroscopic redshift follow the same trend of those with only a
photometric redshift.
Similar trends are observed for the
other 3 IRAC channels (3.6\,$\mu$m, 4.5\,$\mu$m and 8.0\,$\mu$m, not shown here). 
In principle, redder colors could suggest higher 
than average redshifts due to $K$-correction effects, but this is
ruled out by the weak trend with redshift that we observe, e.g., for the ($K-5.8\mu$m) color.
We conclude that the optical to near-IR rest-frame colors of mid-IR
excess sources are intrinsically redder than that of normal sources.
These results imply that whatever the cause of
\24mu    flux density excess in these galaxies, this excess is detectable 
also toward shorter wavelengths as seen in the IRAC bands. 
The distribution in the SFR ratio shown in 
Fig.~\ref{fig:SFR_ch4} for $K-5.8\mu$m$<1$
colors supports our choice of the factor of $\sim3$ ratio 
in Eq.~1 for defining mid-IR excess sources.

\subsection{Median SEDs of galaxies with and without mid-IR excess}

The systematic difference observed for colors reflects different typical
SEDs between the UV/optical and the near-IR rest-frame.
We have derived multiwavelength rest-frame SEDs 
of mid-IR excess and normal galaxies by computing the
median magnitude for galaxies in each sample. Results are shown in Fig.~\ref{fig:SED_N}
for GOODS-N and GOODS-S. This visually confirms that 
excess emission is detected
at all rest-frame wavelengths greater than 1$\mu$m. The typical $SFR_{\rm UV, corr}$
of the 2 samples are very close
for GOODS-N, as can be inferred from the similarity of the SEDs below 1$\mu$m 
[implying also similar median
$E(B-V)$].  Being much deeper in $K$-band, the GOODS-S 
mid-IR-normal sample has lower $SFR_{\rm UV, corr}$
than the mid-IR-excess sample by some 25--30\%. 
Mid-IR excess sources tend to be found more often at brighter $K$-mags,
as one can argue from the figure; see also the next section. 
This explains the larger 
difference in the normalization of the SEDs above 1$\mu$m rest-frame. At 
the shortest wavelengths probed here, it is instead the normal
galaxies that have brighter median magnitudes. 
This is due to the fact that fainter galaxies are bluer (see e.g.
also Daddi et al.~2004b).

\begin{figure}    
\centering 
\includegraphics[width=8.8cm]{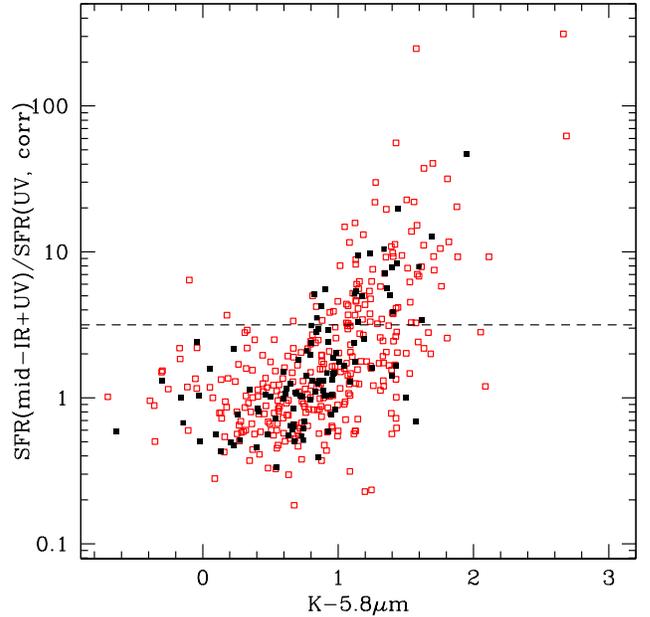}     
\caption{The mid-IR to UV based SFR ratio is plotted as a function of the ($K-5.8\mu$m) color, in the AB scale, for the GOODS-S field. 
Symbols are as in Fig.~\ref{fig:zz_S}. The figure shows that mid-IR excess 
sources are systematically redder in $K-5.8\mu$m color than normal galaxies.
}
\label{fig:SFR_ch4}
\end{figure}

\begin{figure*}    
\centering 
\includegraphics[width=18cm]{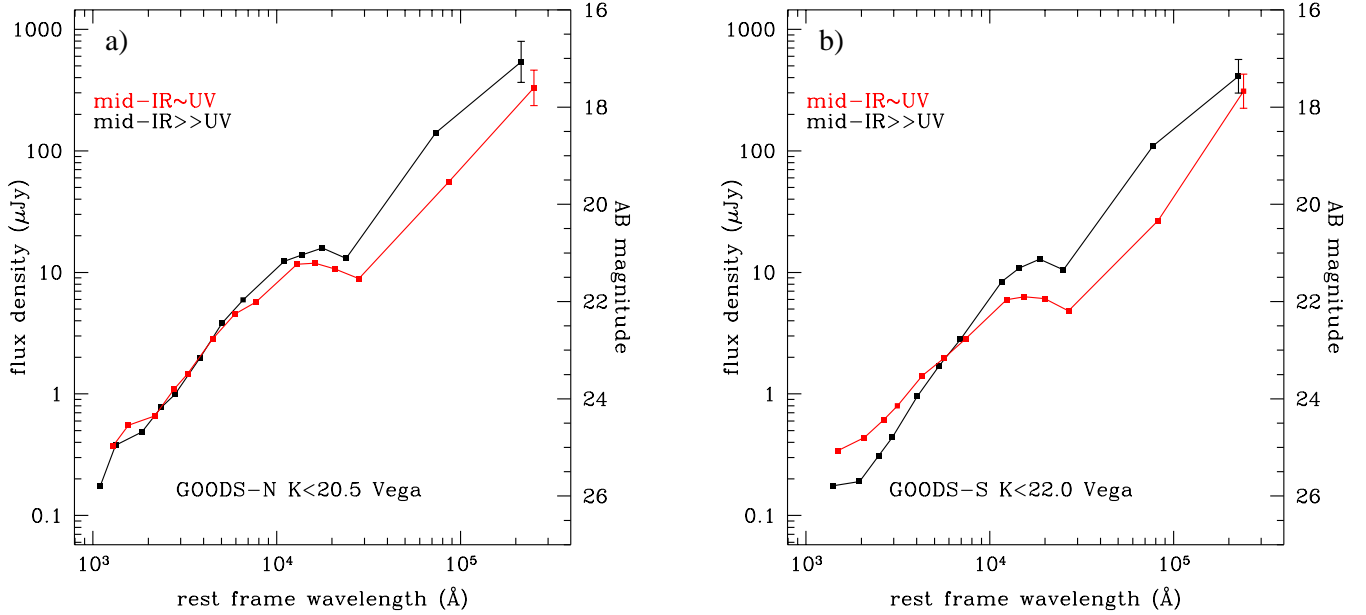}
\caption{Median SEDs of mid-IR excess objects (black) 
compared to those of galaxies with more normal mid-IR
properties (red). Left panel (a) is for the GOODS-N field, limited to $K<20.5$; right panel (b) is for GOODS-S, to $K<22$. 
The GOODS-S photometry includes $BVIzJHK$, the 4 {\it Spitzer}~IRAC bands, and {\it Spitzer}~MIPS \24mu    and 70\,$\mu$m. In GOODS-N 
we have the same, plus U-band.
}
\label{fig:SED_N}
\end{figure*}

More importantly, we notice that the peak of the stellar dominated component
of the SED (say, below 2-3$\mu$m rest-frame) of mid-IR
excess objects is located at longer wavelengths than in normal galaxies (e.g., IRAC $5.8\mu$m band versus $4.5\mu$m band), where
the peak is consistently located at $\sim 1.6\mu$m as expected 
(e.g. Sawicki 2002).
This difference persists when accounting for the slightly higher redshifts 
of mid-IR excess objects ($z\sim2.1$
versus $z\sim1.9$ on average). 
In order for this difference to vanish, one might assume that mid-IR excess
objects are at even higher redshifts than inferred here.  
In turn, the mid-IR excess would vanish if we assume that these galaxies 
are actually systematically at lower redshifts than what estimated here
(which would increase more
strongly $SFR_{mid-IR+UV}$ than $SFR_{\rm UV, corr}$). 
However there is no way to
eliminate both the
mid-IR rest-frame excess and near-IR rest-frame excess by simply
moving the SEDs in 
redshift.\footnote{If these were otherwise normal galaxies 
(also in their mid-IR) with 
just photometric redshifts
(and thus $SFR_{mid-IR+UV}$) largely overestimated, we should 
have found the peak of the
stellar SEDs at lower rest-frame wavelengths than in 
normal galaxies. 
If we had instead underestimated the redshifts of 
the sources (but notice that objects with $z_{spec}$ and $z_{phot}$ behave 
consistently in Fig.~\ref{fig:SFR_ch4}), 
placing them at even higher redshifts
would further increase the strength of the 
excess seen at \24mu.} 
This is an important piece of evidence demonstrating that we are
finding mid-IR excess
galaxies not simply as the result of
biases in the photometric redshifts.

In principle, 
an SED peak longward of $1.6\mu$m could be due to extremely strong dust 
reddening (e.g., Chary et al. 2007). 
However, this is disfavored by the comparison of the UV slopes, and
further by the agreement between the radio and UV SFRs 
which rules out substantial additional obscuration 
in the UV (beyond the obscuration that we estimate based on the UV slopes).
We also tested that these results are robust if restricted to sources with 
spectroscopic redshifts only, and if
excluding (in GOODS-N) sources that are radio detected and could be 
truly UV obscured cases. 
We conclude that from the SEDs there is compelling evidence that the mid-IR 
excess is intrinsic and extends
to near-IR rest-frame wavelengths, in such a way that the SED shape is 
altered and its peak moved to 
wavelengths longer than 1.6$\mu$m (to about 1.8-1.9$\mu$m). 
The required excess component increases sharply with wavelength
up to $\sim10\mu$m rest-frame.

Stacking of sources at 70\,$\mu$m (after excluding the few individual detections)
yields $\simgt3 \sigma$ detections for both mid-IR excess and normal
galaxies in both GOODS-N and GOODS-S fields (Fig.~\ref{fig:SED_N}). 
In both fields, mid-IR excess
and normal galaxies have 70\,$\mu$m stacked flux densities in agreement within the 
errors, although in both fields the mid-IR excess sources are marginally brighter
(less than $1\sigma$ significance in both cases). This is consistent with
the two classes having comparable underlying SFRs.  This also implies
that the excess emission is getting less important, in relative terms, when
sampling $\sim20\mu$m rest-frame wavelengths with the 70\,$\mu$m imaging.
The observed 70\,$\mu$m to \24mu    ratio is 3.2 and 3 for mid-IR excess galaxies in
GOODS-N and GOODS-S
respectively, less than half the ratio expected for a local ULIRG if placed
at $z\sim2.1$. The ratios for the normal galaxies are in better agreement with those
expected from the Chary \& Elbaz (2001) models based on local correlations.
This trend is similar to what found by Papovich et al. (2007).

The mid-IR excess therefore manifests itself as a relatively warmer continuum 
dust emission, 
with respect to the star formation components, whose contribution peaks
in the mid-IR but it is still detectable in the near-IR rest-frame. 
This kind of warm emission is unlikely  to be star formation dominated,
as the reprocessed emission from star formation declines very steeply below 
$5\mu m$ (Genzel et al 1998; Laurent et al 2000), and is a clear signature
of the presence of an AGN. 
These galaxies could be similar
to those selected as putative AGN hosts on the basis of
power-law continua over the same wavelength range
(e.g., Alonso Herrero et al.~2006; Donley et al.~2007),
but with a smaller relative contribution of the AGN at least in the IRAC 
bands, at near-IR rest-frame wavelengths.
By comparing the excess at near-IR and mid-IR rest-frame
wavelengths, we find that
the additional contribution has a shape comprised
between that of the nucleus of
NGC~1068 and of Mrk~231 (at a fixed \24mu    excess, NGC~1068 underpredicts 
and Mrk~231 overpredicts the median excess over the IRAC bands).

This contamination observed in the IRAC bands will have consequences for
SED fitting with stellar population models to derive 
stellar population parameters and stellar masses.
We explore this issue in more 
detail in Maraston et al.~(in preparation).
This contamination can also affect the estimation of photometric 
redshifts of $z\sim2$ galaxies when using IRAC bands, as this estimate
relies also on the location of the 1.6$\mu$m bump of the stellar SEDs.
Comparing to our set of spectroscopic redshifts, we can in fact detect
a small bias, with the photometric redshifts of mid-IR excess galaxies
being slightly overestimated.
Compared to normal galaxies, the median difference between photometric 
and spectroscopic redshifts is larger by about 0.04 (0.09) in GOODS-S 
(GOODS-N), just a few percent of $(1+z)$.

It is also important to recall here that the warm continuum contamination of 
the  SEDs disappear, or at least is entirely negligible, over the UV rest 
frame. This is based on the median SEDs shown, and also on  the
consistency of UV based SFR estimates and submm, radio,
and 70\,$\mu$m, for both mid-IR excess and normal galaxies, 
as extensively discussed in \hbox{Paper~I}. 
In addition, no AGN emission lines are generally detectable
in the UV from the mid-IR excess objects, even in the ultradeep 15--30 
hour integration
GMASS spectroscopy. This is also
consistent with the result that the rest-frame UV morphological properties 
of mid-IR excess and normal galaxies are similar.
\begin{figure*}    
\centering 
\includegraphics[width=18cm]{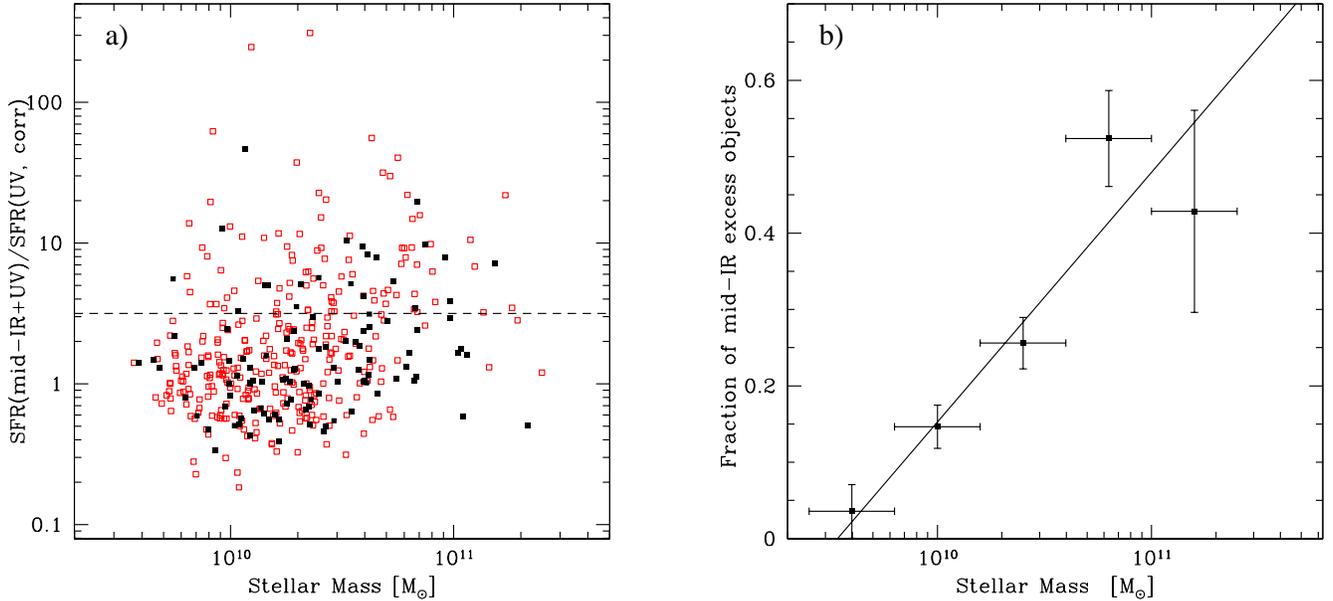}       
\caption{Left panel (a):
the ratio of mid-IR versus UV inferred SFR is plotted against stellar mass. Colors and symbols are as in Fig.~\ref{fig:zz_S}.
Right panel (b): 
the fraction of mid-IR excess objects inside star forming galaxies 
as a function of stellar mass. 
Bins are spaced by 0.4~dex and contain sources within $\pm0.2$~dex in stellar mass (i.e., the 5 bins plotted
are independent).
The line is a weighted linear fit to the data. Error bars on the 
fractions are Poisson, while error bars in the x-axis reflect 
the width of the bin.
}
\label{fig:EXC_Mass}
\end{figure*}

\subsection{Mid-IR excess and  stellar mass} 

We have checked if and how the incidence of mid-IR excess galaxies varies
with the inferred stellar mass (Fig.~\ref{fig:EXC_Mass}).
Stellar masses for our sample are estimated from the empirically calibrated
relations presented in Daddi et al.~(2004b), as discussed in \hbox{Paper~I},
based on the observed photometry in the $B$-, $z$- and $K$- bands.
We find that mid-IR excess sources tend to preferentially reside in
galaxies with larger than average stellar masses. The fraction of mid-IR
excess sources increases steeply with the mass of the galaxies.

We have investigated if the relation shown in Fig.~\ref{fig:EXC_Mass} could be the result of a selection
effect.  First, we have checked that the trend formally
remains if we add to the 
sample the galaxies with uncertain
measurements of the UV slope (and therefore of the UV SFR and mid-IR excess). 
Also, the sample plotted in Fig.~\ref{fig:EXC_Mass} is not mass
complete to the lowest masses it
formally reaches ($\sim0.5\times10^{10}M_\odot$). Given our $K$-band 
selection to $K<22$ (for the GOODS-South sample shown in the figure
on which this analysis is based), we expect to be largely complete 
only above $\sim(2$--$3)\times10^{10}M_\odot$.   
We would miss star forming galaxies below this stellar mass threshold
if they have very red colors. However, we have verified
that we still find statistical evidence for an increasing fraction of mid-IR 
excess sources with stellar mass if we limit the sample to the range
in stellar masses that is expected to be highly complete. 
Similarly, if we limit our analysis to blue galaxies only, for which we are mass complete
over nearly the whole mass range, we still recover the trend 
shown in Fig.~\ref{fig:EXC_Mass}.

The {\em warm continuum} contribution detected over the IRAC bands is 
unlikely to bias the results on the dependence of the fraction of mid-IR excess
galaxies with stellar mass. 
Already over the IRAC bands the typical
excess contribution is less than 0.5 mags\footnote{This is true for
both GOODS fields;  
we recall that the larger difference in the SED normalizations over the IRAC bands that is
observed in GOODS-S is mainly due to 
the different median stellar mass between mid-IR excess and normal galaxies
(Fig.~\ref{fig:EXC_Mass}).} 
and we recall that the stellar
mass estimates are not derived using the IRAC bands, but only up to $K$-band.
The presence of a few tenths of mags 
of AGN contribution in the observed $K$-band
would not alter substantially the trend.
Therefore we conclude that this result is solid and not simply a selection bias.

In \hbox{Paper~I}, we describe a strong correlation between
stellar mass and $SFR_{\rm UV, corr}$.
Because of this correlation, the fraction of mid-IR excess
objects also increases with
$SFR_{\rm UV, corr}$. This effect
is less pronounced and less evident than the one found for the stellar mass.

\begin{deluxetable*}{lccccccccccc}
\tabletypesize{\scriptsize}
\tablecaption{X-RAY STACKING ANALYSIS OF GOODS $z=2$ GALAXIES}
\tablewidth{0pt}
\tablehead{
\colhead{SAMPLE} &
\colhead{$N$}&
\colhead{Exp}&
\colhead{Counts}&
\colhead{Counts}&
\colhead{Counts}&
\colhead{S/N}&
\colhead{S/N}&
\colhead{S/N}&
\colhead{Flux}&
\colhead{Flux}&
\colhead{Flux} \\
\colhead{} &
\colhead{}&
\colhead{SB}&
\colhead{FB}&
\colhead{SB}&
\colhead{HB}&
\colhead{FB}&
\colhead{SB}&
\colhead{HB}&
\colhead{FB}&
\colhead{SB}&
\colhead{HB}\\ 
\colhead{} &
\colhead{}&
\colhead{Ms}&
\colhead{$10^{-6}$~s$^{-1}$}&
\colhead{$10^{-6}$~s$^{-1}$}&
\colhead{$10^{-6}$~s$^{-1}$}&
\colhead{}&
\colhead{}&
\colhead{}&
\colhead{$10^{-17}$~cgs}&
\colhead{$10^{-17}$~cgs}&
\colhead{$10^{-17}$~cgs} 
}
\startdata
NORMAL: & & & & & & & & & & &\\
 & & & & & & & & & & &\\
$K<20.5$ N &69 &	121.2 &	2.05 &1.42 &0.65 &6.6 &8.4 &2.4 &1.78 & 0.74 &1.32\\
$K<20.5$ S &58 &	49.3 &	3.33 &2.35 &1.01 &6.8 &8.4 &2.5 &	2.94 &	1.25 &	2.24 	 \\
$K<22.0$ S &175&146.7 &	2.52 &1.86 &0.68 &9.0 &11.7 &2.9 &	2.23 &	0.99 &	1.50 \\     
 & & & & & & & & & & &\\
EXCESS: & & & & & & & & & & &\\
 & & & & & & & & & & &\\
$K<20.5$ N &27 &	47.2 &	3.80 &2.57 &1.26 &7.3 &9.0 &2.8 &3.29 &1.35 &2.55 \\
$K<20.5$ S &27 &	22.8 &	4.09 &2.88 &1.24 &6.0 &7.5 &2.2 &3.61 &1.53 &2.76 \\
$K<22.0$ S & 59 &	49.8 & 	4.01& 	2.13& 1.92& 8.6& 8.0& 5.0& 6.84& 1.07 &5.62$^*$	
\enddata
\tablecomments{
$N$ is the number of sources that were stacked. FB is full band (0.5-8 keV), 
SB is soft band (0.5-2 keV), and HB is hard band (2-8 keV).
10000 Monte-Carlo trials are performed to estimate the
background and S/N for each sample.
Effective exposure time is measured in the soft band.
Fluxes are calculated using the X-ray spectral slope of $\Gamma=0.8$
for the $K<22$ GOODS-S
mid-IR excess galaxies and assuming $\Gamma=1.9$ for the samples that are
not detected in the hard band.\\
$^*$ using the model shown in Fig.~\ref{fig:Xray_flux} we derive a $\sim1.7\times$ higher flux of $9.55\times10^{-17}$\ergpsec.
}
\end{deluxetable*}

\section{X-ray properties}
\label{sec:X}

The overall properties of mid-IR excess sources, as explored in the
previous sections, point to the presence of a warm mid-IR continuum
component, which could be due to AGN activity.  Since we have removed
from the sample
all X-ray identified AGNs in the 2--8~keV~band, 
any powerful ($L_X \gtrsim 10^{43}$ erg s${}^{-1}$) 
AGN present in
the mid-IR excess objects would have to be heavily obscured in
order to escape being individually detected in the X-rays
(note that the observed 2--8~keV band corresponds to
the rest-frame 6--24~keV band).
We can search
for the signature of heavily obscured AGNs below the X-ray
detection limit using X-ray stacking analyses (e.g., Brandt et
al.~2001). Here we use the code adopted by Worsley et al.~(2005), which stacks
sources and calculates the significance of the stacked result using
10,000 Monte Carlo trials. In all cases, we limit the X-ray stacking to
sources within 5.5 arcmin of the {\it Chandra} aim point, to maximize
sensitivity.

\subsection{X-ray stacking results}

\begin{figure}
\centering
\includegraphics[width=8.8cm]{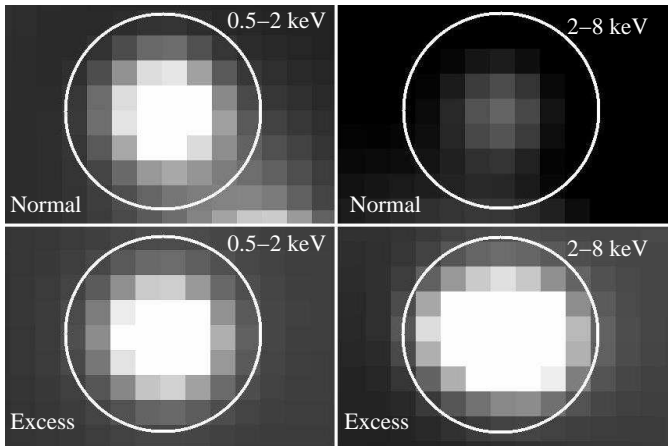}        
\caption{The soft and hard X-ray {\it Chandra} stacked images of normal and
mid-IR excess $z=2$ objects in GOODS-South.  Left panels show the soft
(0.5--2~keV) bands, while the right panels are for hard (2--8~keV)
bands.  The top 2 images are for normal galaxies, while the bottom 2
are mid-IR excess objects.
It is evident that similar soft X-ray fluxes are
detected in the 2 samples, but much stronger hard emission is
detected for the mid-IR excess $z\sim2$ galaxies. 
Images have been smoothed by a 
Gaussian with the size of the PSF. 
The circles (4$''$ diameter) show the expected location of the signal.
}
\label{fig:Xray_dets}
\end{figure}

We stacked 59 mid-IR excess galaxies in GOODS-S and
found a
5$\sigma$ detection in the 2--8~keV band and an 8$\sigma$
detection in the 0.5--2~keV band (Fig.~\ref{fig:Xray_dets}).  The
count ratio (H/S) is $0.90\pm0.13$, 
corresponding to a nominal power law photon index
$\Gamma=0.8^{+0.4}_{-0.3}$. Using $\Gamma=0.8$, we infer an average 2--8~keV 
flux of $5.6\times10^{-17}$~\ergpsec~cm${}^{-2}$ and an average
0.5--2~keV flux of $1.1\times10^{-17}$~\ergpsec~cm${}^{-2}$.  
The flat X-ray spectral slope strongly argues for the
presence of heavily obscured AGN (e.g.,\ see Fig.~2 of Alexander et~al.
2005b).
The 175 {\it normal} GOODS-S sources (i.e., those without
mid-IR excess) give a stacked 3$\sigma$ signal in the 2--8~keV band and
a 12$\sigma$ detection in the 0.5--2~keV band
(Fig.~\ref{fig:Xray_dets}).  The band ratio is 0.36, corresponding to
a spectral index $\Gamma \sim1.7$, which is fully consistent with star
formation dominating the X-ray emission (e.g., Persic \& Raphaeli 2002). 
Using this value of $\Gamma$
we derive a 2--8~keV flux of $1.6\times10^{-17}$~\ergpsec~cm${}^{-2}$
and a 0.5--2~keV flux of $1.0\times10^{-17}$~\ergpsec~cm${}^{-2}$.
Stacking analysis of the smaller GOODS-N sample, that is limited to
$K<20.5$, gives a slightly less secure detection and lower S/N
ratios, although still with hard X-ray fluxes twice as large
for mid-IR excess
versus normal galaxies. Similar properties are seen in the GOODS-S
sample when limited to $K<20.5$.  In the following we concentrate on
the GOODS-S sample to $K<22$, because it has the highest S/N X-ray detections. 
The results of the X-ray stacking are listed in Table 1.

The observed X-ray fluxes can be directly used to estimate luminosities, without
accounting for obscuration.
For the mid-IR normal galaxies in the GOODS-S sample
($<z>=1.9$), we infer isotropic
luminosities of $2.7\times10^{41}$~\ergpsec\ in the 2--8~keV
rest-frame band (mapping very closely to the 0.5--2~keV observed band)
and of $4.2\times10^{41}$~\ergpsec\ in the 5.8--23.2~keV band
rest-frame, assuming a $\Gamma=1.9$ spectral slope. 
For mid-IR excess galaxies in GOODS-S ($<z>=2.1$) the fluxes
correspond to X-ray luminosities of $3.8\times10^{41}$~\ergpsec\ in
the 2--8~keV rest-frame band and $2\times10^{42}$~\ergpsec\ in the
6.2--24.8~keV band rest-frame, for the spectral slope of $\Gamma=0.8$ derived
from the count ratios.

\begin{figure}
\centering
\includegraphics[width=8.8cm]{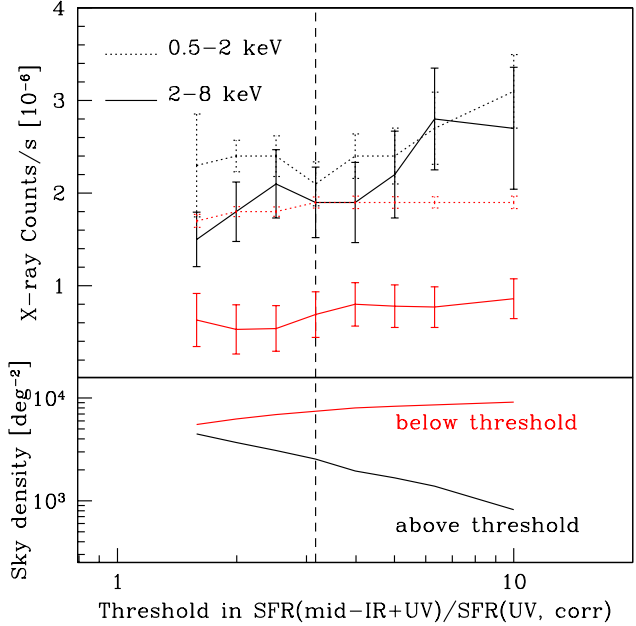}         
\caption{(Top panel): X-ray counts from stacking of mid-IR excess
versus mid-IR normal sources for a variety of thresholds separating
these two classes: from 0.2 to 1.0 dex in the ratio of
$SFR({\rm mid-IR+UV})/SFR({\rm UV, corr})$.  Black lines are for mid-IR excess
sources (with ratio above the threshold), while red lines are for
mid-IR normal sources (with ratio below the threshold).  Solid lines
show hard (2--8~keV band) counts, while dotted lines show soft 
(0.5--2~keV band) counts, from stacking {\it Chandra} data in the 1Ms GOODS-S
field. (Bottom panel): the sky density of the two populations, as a
function of the defining threshold. We are not applying here completeness
corrections to the sample, see Sect.~5.1.}
\label{fig:Xray_Thresh}
\end{figure}

We have explored how these X-ray stacking results depend on the
threshold of 0.5~dex in the ratio of $SFR({\rm mid-IR+UV})$ to $SFR({\rm UV, corr})$
that we have adopted in Eq.~1 for defining mid-IR excess
sources. We have varied the threshold from 0.2~dex to 1.0~dex, and
performed X-ray stacking separately for sources above and below the
threshold.  Fig.~\ref{fig:Xray_Thresh} shows that the general X-ray
properties of the "normal" populations are quite insensitive to the
threshold at which the {\em excess} sources are identified, given that
the latter ones are in all cases a minority. For the mid-IR excess sources,
the X-ray stacking confirms that a threshold 2--5\ in
the SFR ratio effectively singles out the population of hard
X-ray emitters (with similar count rates in both hard and soft {\it Chandra}
bands). The X-ray count rates, and therefore the average luminosities of
the mid-IR excess sources, increase with the threshold, 
while their space density
obviously decreases, but the accuracy in the
measurements declines due to lower-number statistics.

A possible concern with stacking analyses is whether a 
small subset of bright sources dominate the
stacked results. We verified that this is not the case by performing stacking
in galaxy subsamples. The results shown in Fig.~\ref{fig:Xray_Thresh} argue
that our X-ray stacking conclusions are robust to the exact definition of the sample.

\begin{figure*}
\centering
\includegraphics[width=17cm,angle=0]{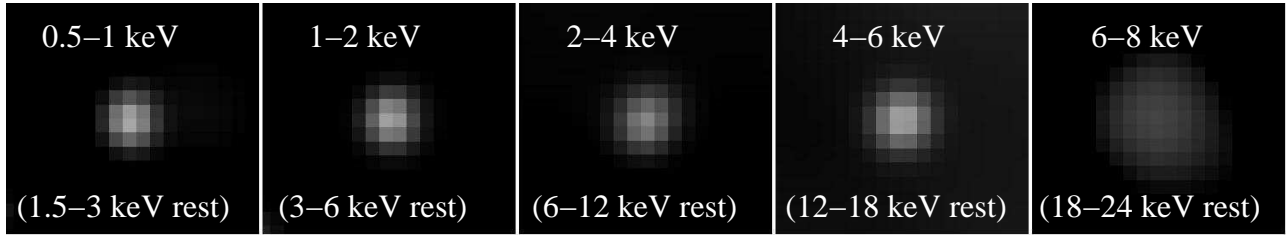}           
\caption{{\it Chandra} sub-band stacking detections of mid-IR excess $z\sim2$
galaxies. The images have been smoothed with the
PSF of the band, to enhance the signal visibility. Each panel size is 12$''$.
}
\label{fig:Subband_CXO}
\end{figure*}

\begin{figure*}
\centering
\includegraphics[width=18cm]{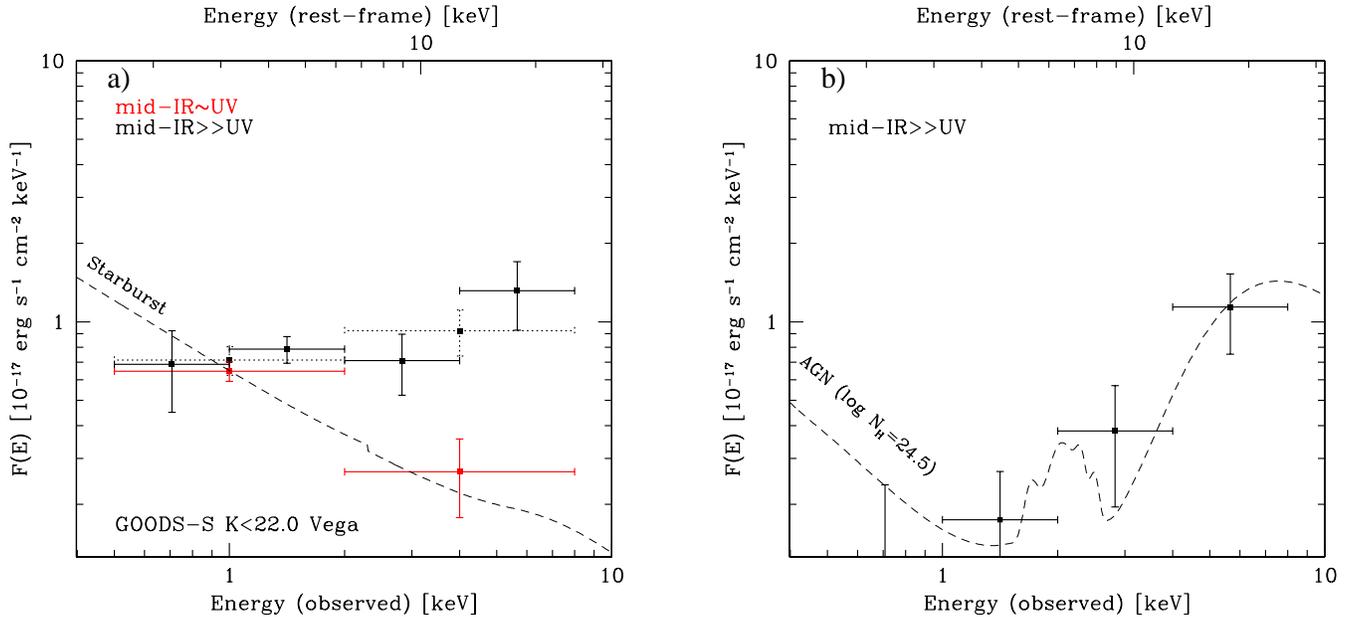}    
\caption{Left Panel (a): X-ray measurements of flux per unit energy
($F(E)$) for mid-IR excess sources (black) and normal sources (red).
It is evident that mid-IR excess sources are much harder,
inconsistent with pure star-formation as indicated by the 
photon index $\Gamma=1.9$ power-law
(Starburst). Sub-band stacking of mid-IR excess sources
(black points with solid error bars) 
reveals that their  $F(E)$ continue to rise at the
highest energies. These sources are consistent with a combination
of star-formation plus a heavily obscured AGN component
rising toward the highest energies.  Right Panel (b): the result of
subtracting the star-formation rate component from the sub-band
{\it Chandra} stacking of the mid-IR excess galaxies, normalized to the soft
band point (allowing for a plausible range of uncertainty to this
quantity does not substantially alter the results). The overplotted
line is an AGN model with $N_{\rm H}=10^{24.5}$~cm${}^{-2}$
(Gilli et al.~2007) convolved with the redshift distribution of the
sample.  The bump in the models 
near 2~keV (in the observed frame) is due to the
strong Fe emission line expected to be prominent in obscured AGNs.  }
\label{fig:Xray_flux}
\end{figure*}

Although the X-ray band ratio indicates the presence of obscured AGNs,
it does not fully characterize the properties of the sources.  We can
gain additional insight into the X-ray spectral properties of mid-IR
excess sources by stacking the data inside narrower energy bands.
Stacking the GOODS-S data using the sub bands 0.5--2~keV, 1--2~keV, 2--4~keV, 4--6~keV, 4--8~keV, and 6--8~keV, one obtains the results
shown in Figs.~\ref{fig:Subband_CXO} and~\ref{fig:Xray_flux} and Table
2.  Significant ($\simgt 3\sigma$) detections are found in all of the
sub bands, except for the 6--8~keV band where the detection is only at
the 1.8$\sigma$ level (still noteworthy given the low sensitivity of
{\it Chandra} at these high energies).  In summary, the mid-IR excess
sources exhibit a hard X-ray spectrum which becomes harder with
increasing energy, an unequivocal signature of highly obscured AGN
activity.

\subsection{X-ray inferred SFRs}

The soft X-ray (0.5--2~keV band) fluxes of the samples (mapping
quite closely to the
rest-frame 2--8~keV luminosities at $z=2$) can be used to
estimate SFRs, since for star forming galaxies this is proportional to the
population of massive X-ray binaries
(Ranalli et al 2003; Persic et al 2004; Hornschemeier et al 2005).
The rest-frame 2--8~keV
luminosity of mid-IR excess sources is $\sim 30\%$ higher than that of
mid-IR normal sources, and would thus suggest $\sim30$\% higher SFRs
for the formers. This is in excellent agreement with the
estimate obtained from the UV, where we find the same offset
between the average $SFR_{\rm UV, corr}$ of the two samples and in the
same direction.  Interestingly, the proportionality that we derive is

\beq
SFR_{{\rm UV,corr}}\; [M_\odot yr^{-1}]=2.25\times10^{-40}L_{2-8~keV}\; [erg\;
s^{-1}]
\eeq
\par

\noindent
which is very close to the relation given in Ranalli et al.~(2003)
(the Persic et al. 2004 relation only accounts
for the contribution of high mass X-ray binaries).
The agreement between the UV and X-ray estimated SFRs, for both samples of
normal and mid-IR excess galaxies in GOODS-S,
suggests that no substantial AGN contribution is detected in the
observed 0.5--2~keV band of these mid-IR excess objects.
Similarly, we find that the soft X-ray fluxes in the mid-IR
excess samples in both fields, when limited to $K<20.5$, are fully accounted
for by the SFR inferred from UV.

\subsection{Evidence for Compton thick AGNs}

Estimating the obscuring column density for the AGNs inside mid-IR
excess sources is crucial for deriving their intrinsic
properties. Their stacked {\it Chandra} 0.5--2 to 2--8~keV band ratio implies an
absorbing gas column density $N_{\rm H}\sim10^{23.3}$~cm${}^{-2}$, for
an underlying standard AGN power-law continuum with $\Gamma=1.9$.
Therefore, these AGNs are clearly obscured, but this estimate of $N_{\rm
H}$ should be regarded as a strict lower limit because star-formation
is contributing to the emission, especially in the soft band.  The
0.5--2~keV emission of the mid-IR excess
galaxies can be entirely ascribed to their ongoing star
formation, as discussed in the previous section. If we consider that
the AGN component in the 0.5--2~keV band is consistent with zero given
the errors, we formally obtain a 2$\sigma$ lower limit 
$N_{\rm H}>10^{23.9}$~cm${}^{-2}$ for an intrinsic AGN power law continuum with
$\Gamma=1.9$.

We further used the result of spectral stacking shown in Table 2 and
Fig.~\ref{fig:Xray_flux} to improve our constraints on $N_{\rm H}$.
The right panel of Fig.~\ref{fig:Xray_flux} shows the X-ray spectrum
at 2--20~keV rest-frame energies of the mid-IR excess sources, after
subtraction of a star-formation component (dashed line model shown in
left panel of Fig.~\ref{fig:Xray_flux}). We have fitted this AGN
component with the Gilli, Comastri \& Hasinger (2007) models for
obscured AGNs, where $N_{\rm H}$ is increased in steps of 1 dex up
to $N_{\rm H} = 10^{25.5}$~cm${}^{-2}$ (this approach is admittedly
crude, but there is little point for a more refined approach in this work).
At the highest levels of obscuration, beyond the Compton thick
regime of $N_{\rm H} = 10^{24}$~cm${}^{-2}$, reflection components start
to be important in the AGN emission
in the rest-frame $\sim$2--8~keV (observed {\it Chandra}
0.5--2~keV for $z\sim2$), as well as the emission from the Fe $K\alpha$
line at 6.4~keV, and both are properly included in the models of Gilli
et al.~(2007). Having convolved the templates with the observed
redshift distribution of our sample, the best fit is obtained for
$N_{\rm H} = 10^{24.5}$~cm${}^{-2}$, with both cases $N_{\rm H} =
10^{23.5}$~cm${}^{-2}$ and $N_{\rm H} = 10^{25.5}$~cm${}^{-2}$ producing a
worse fit with $\Delta\chi^2>5.8$. According to Avni et al.~(1976),
this implies that they can be rejected at the $>98.5$\% confidence
level.  For the best fitting case of $N_{\rm H} =
10^{24.5}$~cm${}^{-2}$, the observed 2--8~keV band would be obscured by a
factor of $\sim4$, with respect to the intrinsic, unobscured emission.

\begin{deluxetable}{lccc}
\tabletypesize{\scriptsize}
\tablecaption{STACKED X-RAY SPECTRA}
\tablewidth{0pt}
\tablehead{
\colhead{BAND} & \colhead{Counts} & \colhead{S/N} & \colhead{Flux}\\
\colhead{} & \colhead{$10^{-6}$~s$^{-1}$} & \colhead{} & \colhead{$10^{-17}$~cgs}
}
\startdata
0.5-1keV &	0.57 &	2.9 &	0.34 \\
1-2keV	 &	1.67 &	8.5 &	0.79 \\
2-4keV	 &	0.89 &	3.8 &	1.42 \\
4-8keV	 &	1.16 &	3.4 &	5.25 \\
6-8keV	 &	0.47 &	1.8 &	4.94 
\enddata
\tablecomments{Results of
the X-ray spectral stacking for  mid-IR excess objects in GOODS-S to $K<22$.
Fluxes calculated using the overall X-ray spectral slope of $\Gamma=0.8$,
estimated from the X-ray band ratio.
S/N of the data calculated using the background estimated from 10,000 MC trials.
}
\end{deluxetable}

Therefore, the X-ray spectrum of mid-IR excess sources is fully
consistent with the presence of two major components: a star-formation
component (modeled here with a power-law emission of $\Gamma=1.9$),
plus a highly obscured AGN component with (average)
$N_{\rm H}\sim10^{24.5}$~cm${}^{-2}$, i.e., mildly Compton thick.

\subsection{Column density and mid-IR luminosities}

The column density of obscuring gas can be independently estimated by
exploiting the correlations between mid-IR and X-ray luminosities for
AGNs, in the assumption that the mid-IR excess is entirely due to the
AGN.  The mid-IR excess sources to $K=22$ in GOODS-S have average
(median) \24mu    flux densities of 145 (131) $\mu$Jy, while normal sources have
average (median) \24mu    flux densities of 41 (27) $\mu$Jy, consistent with
the ongoing SFR on the basis of local correlations.  We have computed
for all galaxies the \24mu    flux densities expected for their $SFR({\rm UV, corr})$
on the basis of Chary \& Elbaz (2001) models, and compared to the
observed \24mu    flux densities. We find also in this way that for mid-IR excess
galaxies the average (median) excess flux density is 100$\mu$Jy
(80$\mu$Jy). The ratio of \24mu    excess flux density to the total \24mu    flux  density
is  found to be 69\% (68\%) on average (median).  
Therefore, we infer that the typical
excess flux density is $\approx 80$--$100\;\mu$Jy and that the AGN is
contributing typically some 2/3 of the observed \24mu    flux density in these
mid-IR excess galaxies.  These figures imply a ratio of \24mu    flux density to hard
X-ray flux of $\sim2\times10^{15}$ in the [mJy]/[\ergpsec~cm${}^{-2}$]
units used by Rigby et al.~(2004). According to Rigby et al. (2004),
based on the local AGN templates from Silva, Maiolino \& Granato (2004) with a range
of obscuration, such large ratios at $z=2$ would imply column densities
$N_{\rm H}>>10^{24}$~cm${}^{-2}$, or substantially Compton thick
sources. Using the observed correlations between the AGN mid-IR light
and unobscured hard X-ray light from Lutz et al.~(2004), which appear
to hold independently of obscuration at least to $N_{\rm
H}\sim10^{25}$~cm${}^{-2}$, we estimate that unobscured AGNs would
contribute only $\approx 3$--10$\mu$Jy at \24mu    for $z=2$ given
their 2--8~keV X-ray fluxes.  In order to match the Lutz et al.~(2004)
correlation, the 2--8~keV observed flux requires a correction factor of
$\approx 10$--15, i.e., higher than the factor $\sim 4$ derived in the
previous section, and potentially implying Compton-thick AGN with
$N_{\rm H}>10^{25}$~cm${}^{-2}$. 

It is instructive also to compare to the properties of NGC~1068, an
extremely obscured local starburst harboring a heavily Compton thick
AGN with $N_{\rm H}>10^{25}$~cm${}^{-2}$.  If placed at $z=2$, the
NGC~1068 nucleus would have a hard band X-ray flux of
$7\times10^{-18}$~\ergpsec~cm${}^{-2}$ and a \24mu    flux density of
25$\mu$Jy. This would give a mid-IR to X-ray ratio of
$\sim3.5\times10^{15}$ ([mJy]/[\ergpsec~cm${}^{-2}$]) fully comparable
(slightly larger) than that of the mid-IR excess $z=2$
galaxies. Therefore, these $z=2$ sources are consistent with being the
high-redshift analogs of NGC~1068, with overall luminosity scaled up
by a factor $\sim 5$. The AGN emission of NGC~1068 is thought to be
obscured by over a factor of 10, even in the 6--24~keV band (rest
frame).

We conclude from the mid-IR to X-ray comparisons that 
the majority of AGNs shrouded
inside massive mid-IR excess $z=2$ galaxies 
are likely Compton thick, with average
$N_{\rm H}>10^{24}$~cm${}^{-2}$, perhaps even in excess of the estimate of
$N_{\rm H}\sim 10^{24.5}$~cm${}^{-2}$ derived solely from the X-ray stacking,
and possibly in some cases they have extreme obscuration with
$N_{\rm H}\simgt10^{25}$~cm${}^{-2}$.

\subsection{Intrinsic AGN luminosities}

A determination of the intrinsic X-ray luminosity is critical 
to assess the general implications of our findings,
but unfortunately it
is made difficult because the obscuration appears to be very
high.  For the best fitting Gilli et al.~(2007) mildly Compton thick
model having $N_{\rm H}=10^{24.5}$~cm${}^{-2}$, 
the {\it observed} 2--8~keV band is obscured by a factor of $\sim4$.
Consistently using this model to derive
2--8~keV flux from the observed counts yields a flux larger by a factor of 1.7
than that reported in Table~1, computed for $\Gamma=0.8$.
Correspondingly, we estimate that the average, isotropic, unobscured
2--8~keV {\em rest-frame} luminosity of mid-IR excess galaxies is $L_{2-8
{\rm keV}} \sim 10^{43}$~\ergpsec.  
To give an idea of the
uncertainty in the correction for obscuration, we find that for the Gilli
et al. (2007) templates  the
luminosities would decrease by a factor of 2.9 for
$N_{\rm H}=10^{23.5}$~cm${}^{-2}$ and would increase by a factor of 3.5 for
$N_{\rm H}=10^{25.5}$~cm${}^{-2}$ (although we emphasize that the absorption 
correction for the case of $N_{\rm H}=10^{25.5}$~cm${}^{-2}$ is highly
dependent on the assumed reflection efficiency).

As a comparison, we can use the Lutz et al.~(2004) correlation between
unobscured X-ray luminosities and $6\mu$m luminosities for AGNs, which
is found to be independent of obscuration and has a scatter for
individual sources of about a factor of 3.  
The relationship between the
X-ray and mid-IR emission is strongly dependent on the dust-covering
factor around the central source and hence a source with a large dust
covering factor will be more mid-IR luminous than one with a smaller dust
covering factor (see Alexander et al. 2005b).
The average \24mu   
excess flux density attributed to the AGN is converted to an average
$8\mu$m luminosity of order $4\times10^{10}L_\odot$ for the ensemble
of mid-IR excess objects. We estimate $\sim70\,$\% this value for the typical 
$6\mu$m luminosity of our
sources (based on typical SEDs, see also Fig.~\ref{fig:SED_N}).  From
Fig.~7 in Lutz et al. (2004) we would then estimate $L_{2-8 {\rm keV}} \sim
4\times10^{43}$~\ergpsec, a factor of 4 larger than
that obtained directly from X-ray.  In summary, plausible values for
the typical rest-frame 2--8~keV unabsorbed luminosities of mid-IR
excess galaxies appear to be within $(1$--$4)\times10^{43}$~\ergpsec.

This result can be cross-checked by looking at the X-ray luminosities 
of {\em individually} hard X-ray detected  AGNs in our GOODS
samples of $BzK$ selected sources with $1.4<z<2.5$. 
If we are finding the 
heavily obscured, Compton thick, counterparts of the already known 
population of $z\sim2$ relatively unobscured sources, we might expect 
that the Compton thick sources should have typical luminosities similar
to the known AGNs, based on AGN unification models 
(Antonucci 1993; Urry \& Padovani 1995). 
In fact, we find that the average 2--8~keV rest-frame luminosity
of {\em individually} hard X-ray detected  AGNs in our $K$-selected
sample is about $L_{2-8 {\rm keV}} \sim 3\times10^{43}$~\ergpsec, 
neglecting obscuration corrections that are likely negligible in most of 
these cases.
The individually X-ray detected $z=2$ AGNs have a sky 
(and thus space) density that is a factor of $\sim3$ 
smaller than that of our mid-IR excess galaxies in GOODS-S.
Given the estimated X-ray luminosities, the mid-IR excess galaxies are also
consistent with the correlation between optical/near-IR and 2--10~keV 
rest-frame luminosities found by Brusa et al. (2005) for distant
AGNs.

One might also hope to constrain the bolometric
luminosities of the AGNs in our sample, although uncertainties in the AGN
bolometric correction factors are currently large 
(see, e.g., discussions in Pozzi et al.~2007 and Hopkins et al.~2007). 
Assuming an AGN bolometric correction
derived from the Elvis et al.~(1994) SED, from the 2--8~keV X-ray
luminosity we estimate
$L_{BOL}^{AGN}\sim(3$--$12)\times10^{44}$~\ergpsec, or about
$(0.8$--$3)\times10^{11}L_\odot$.  The large range reflects only the
uncertainty in the derived 2--8~keV rest-frame luminosity discussed above,
as we are neglecting the unknown uncertainty in the bolometric
correction for these galaxies.  If instead we start from the mid-IR
excess, with a typical luminosity of $10^{11}L_\odot$, the Elvis et al.~(1994) 
SED would predict a bolometric correction of a factor of 10, or
$L_{BOL}^{AGN}\sim10^{12}L_\odot$. Given that our galaxies are heavily
obscured in X-rays to a similar degree, it may be more appropriate
to use the mid-IR bolometric correction for the NGC~1068 nucleus,
which is a factor of 3.  This would lead us to estimate
$L_{BOL}^{AGN}\sim3\times10^{11}L_\odot$.  So, we are left with a
plausible range spanning from 0.8 to $3\times10^{11}L_\odot$, although a
value as high as $10^{12}L_\odot$ cannot be completely excluded at
this stage.  Admittedly, a very large uncertainty affects the
estimated AGN bolometric luminosity.

From the ongoing star formation activity
we estimate $L_{BOL}^{SF}\sim5\times10^{11}L_\odot$ for the typical
mid-IR excess object, comparable to the bolometric luminosity of the AGN.

\section{Implications}
\label{sec:impli}

\subsection{The sky and space density} 

The main result of our study is the
uncovering of a large population of Compton thick AGNs at $z\sim 2$.
We characterize
here the abundance of this population.  The sky density of the mid-IR
excess AGNs in the GOODS-S field is $\sim0.7$~arcmin${}^{-2}$ to
$K<22$ (59 sources in a 5.5$'$ radius region), after correcting for
the 10\% incompleteness in our sample due to the requirement of
non blended IRAC photometry (see \hbox{Paper~I}).  This estimate could
increase by a factor of $\sim $2, accounting for the sources that we
excluded from the analysis not being able to properly measure for those
either  $SFR(mid-IR+UV)$ or $SFR({\rm UV, corr})$ in Eq.~1.  (which requires a
well measured UV slope and \24mu    detection, mainly, see Sect.~2).  As the
excluded sources lacking \24mu    detections are less likely to be
mid-IR excess galaxies, we conservatively
correct the sky densities only by a factor
1.3, to account for the overall fraction of red star forming galaxies
that were
excluded from the analysis 
for not having a well measured UV slope. This implies a sky
density of 0.9 arcmin$^{-2}$, or some 3200~deg$^{-2}$.  Using the
volume within $1.4<z<2.5$ (see Fig.~2 in \hbox{Paper~I}) we infer a
space density of $2.6\times10^{-4}$~Mpc${}^{-3}$ for the adopted
$\Lambda$CDM cosmology.  

We are dealing with detections of {\em
average} properties only (X-ray stacking), and there is likely a
spread in luminosities and $N_{\rm H}$ among the sample.  In principle
only a fraction of the stacked mid-IR excess sources might be contributing
the detected hard X-ray flux, e.g. in the case of a large contamination of
the mid-IR excess galaxy sample from galaxies with peculiar mid-IR SEDs
that are otherwise powered by star formation. 
In that case, the space density of this obscured AGN  population
would be lower by the same fraction, and its average X-ray flux
and luminosity would be proportionally
higher. The contributing fraction cannot be too small, 
i.e. at least $\approx0.5$, otherwise we would have
detected directly most of these sources in our {\it Chandra} data. Also, the
similarity in inferred X-ray luminosities of these $z=2$ Compton thick AGNs 
to individually detected $z=2$ sources (Sect.~4.5) argues against a lower
fraction since this would boost their intrinsic luminosities.
With this
caveat, we have included all mid-IR sources in the computation of the space
density of Compton thick AGN candidates assuming that the mid-IR
excess is due to the AGN as discussed above. Relaxing this hypothesis would
not substantially affect any of the implications that we derive from this work.

\begin{figure}
\centering
\includegraphics[width=8.8cm]{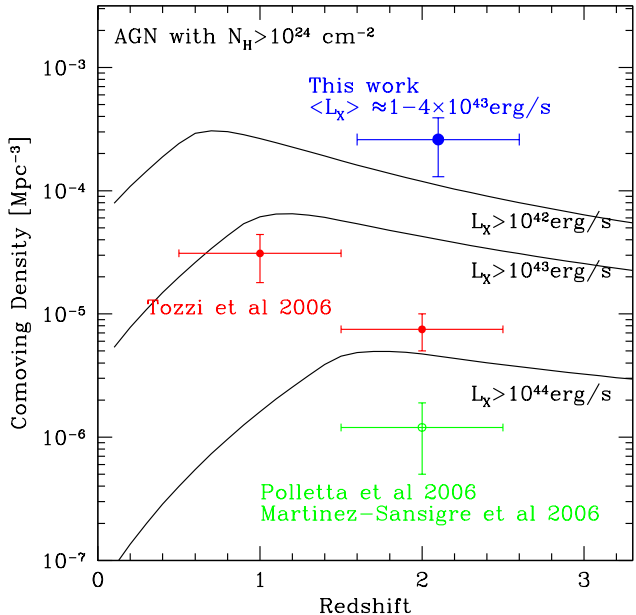}
\caption{The space density of Compton thick AGNs. The blue circle is
our $z=2.1$ estimate,  where we
allow for a factor of 2 uncertainty. The red points are taken from
Tozzi et al.~(2006), based on direct {\it Chandra} detections in GOODS-S. The
green points show the density that we crudely estimate for the survey
of Polletta et al.~(2006) and Martinez-Sansigre et al.~(2006), accounting
in the latter case for completeness correction due to their radio
preselection. The lines
show the predictions of the background synthesis model of Gilli et al~(2007),
as a function of the limiting X-ray luminosity.
}
\label{fig:Xvol}
\end{figure}

\subsection{The predominance of Compton thick AGNs at high redshift}

It is interesting to compare the properties of this previously
undetected population of obscured, Compton thick,
AGNs to AGN populations already known.
Over the same redshift range $1.4\simlt z \simlt 2.5$,
the sky density of mid-IR excess, Compton thick AGNs is a
factor of $\sim2$ higher than that of X-ray sources individually detected in
X-rays in GOODS-S (Zheng et al 2004), a factor
$\approx$~10 higher than X-ray detected
submillimeter-emitting galaxies (Alexander et~al. 2003b), and $\approx 20$
times higher than X-ray undetected power-law AGN candidates (Donley
et al. 2007).

We can assess the obscured-to-unobscured
AGN ratio by comparing to the space density of unobscured AGNs
selected
in the soft band by Hasinger et al.~(2005).
For $L_X>10^{42}$~\ergpsec, they find a space
density of $\sim(3$--$4)\times10^{-5}$ at $z=2$, a factor of $\simgt5$
smaller than what we find for obscured mid-IR excess AGNs at the same
redshift.  This implies that the ratio of obscured to unobscured AGNs
is larger than at least 5, and is probably much larger yet, consistently with
the models expectations (e.g., Gilli et al 2007).
Comparing to the luminosity function for hard-X-ray selected AGNs at
$z=2$ (Ueda et al.~2003; La Franca et al.~2005), the Compton thick sources 
that we
have found have a space density that is larger by a factor of 2--3 than that 
of the sources already known and overall with similar luminosities.
The difference would decrease if it is only a fraction of mid-IR excess sources
that are hosting an X-ray luminous AGN.
A more detailed comparison is
hampered by the uncertainties inherent the detailed properties of our
samples, as discussed in the previous sections. Overall, these results 
suggest that a dominant fraction of X-ray luminous
AGNs are heavily obscured, in general agreement with expectations.

It is also interesting to compare to previously known populations of 
Compton thick AGNs.
Tozzi et al.~(2006) identify 6 Compton thick AGNs with $N_{\rm H}>10^{24}$
at $1.5<z<2.5$, detected by {\it Chandra} in the full GOODS-S, which
translates into a space density at least an order of magnitude lower
than that inferred here for our mid-IR excess objects. A larger
density of Compton thick AGNs is directly detected at $z\sim1$ by
Tozzi et al.~(2006), but this is
still a factor of at least 3 lower than our estimate
at $z=2$ (Fig.~\ref{fig:Xvol}). Overall, the space density of Compton
thick AGNs that we infer at $z=2$ is reasonably consistent, within 
a factor of 2, with the predictions
of the background synthesis model of Gilli et al.~(2007;
Fig.~\ref{fig:Xvol}), when counting all the Compton thick sources 
in the models with
$10^{42}<L_X<10^{44}$~\ergpsec, 
which are found to have an average X-ray luminosity
of $L_X = 1.3\times10^{43}$ that is similar indeed to that of our 
sources.\footnote{We emphasize that the space density of Compton thick AGNs
at $z=2$ in the Gilli et al. 2007 models is constrained by the shape of the
missing X-ray background, and is uncertain by a factor of a few. This is because
$z=2$ Compton thick AGN in the models produce only a small fraction of the 
missing background, see Sect.~5.3.}
This large population of Compton
thick $z\sim2$ AGNs has probably similar intrinsic properties to that
found by Martinez-Sansigre et al.~(2006), who select $z\sim2$ AGNs as
mid-IR bright sources with a radio detection and a faint IRAC
counterpart.  They have uncovered a population of AGNs with a sky
density of $\approx10$~deg${}^{-2}$, turning out to be heavily obscured
sources in nearly all cases.  Their AGNs are likely the tip of the
iceberg of the population found here, given their sky density some 2 orders of
magnitude lower than that of our sources, but much higher bolometric
luminosities.  Similarly, Polletta et al.~(2006) find a population of
luminous Compton thick AGNs at $z\sim2$ using {\it Spitzer}, 
with a sky density of about 25~deg${}^{-2}$.

In parallel to this work, Fiore et al. (2007) detected a similar population
of $z\sim2$ obscured AGNs from X-ray stacking of galaxies with large 24\,$\mu$m
to optical flux ratios. Given their selection criteria, 
we expect that many of the Fiore et al. objects would classify  as mid-IR 
excess galaxies.

\subsection{Contribution to the X-ray background}

Given an observed average hard X-ray flux of
$\sim10^{-16}$~\ergpsec~cm${}^{-2}$ 
(calculated directly from the model in Fig. 9 which gives a factor 
$\sim1.7\times$ higher flux than given in Table 1; see Sect. 4.5)
and the sky density of
3200~deg${}^{-2}$, the mid-IR excess sources produce a background of 
0.7~keV~s${}^{-1}$~cm${}^{-2}$~sr$^{-1}$ in the 2--8~keV band. This is a
quite small contribution to the total background in this band, 
which is not surprising given that this band is almost completely
resolved by direct detections of individual AGNs (at least up to
$\approx$~6~keV; Worsley et al.~2005; 2006).
The relative contribution of these mid-IR excess sources increases towards
higher energies, as these sources are strongly obscured. 

As an
example, we have tried to extrapolate the background flux emitted in
the 10--30~keV band.  Using the best fitting Gilli et al.~(2007) model
with $N_{\rm H}=10^{24.5}$~cm${}^{-2}$, we estimate that these sources
would produce a background flux of about 1.2
keV~s${}^{-1}$~cm${}^{-2}$~sr$^{-1}$ in the 10--30~keV band.  For the
physically motivated SEDs described in Gilli et al.~(2007) this
prediction turns out to be only weakly dependent on the assumed
$N_{\rm H}$, for AGNs around the Compton thick regime. However, for
the X-ray spectral shape at the highest energies of NGC~1068 and
NGC6240 (Vignati et al.~1999), we would derive a contribution a factor
of 2 larger than above, being flatter at the highest energies than the
Gilli et al. (2007) models (on the other hand, e.g., the Circinus
galaxy has a spectral shape even softer than the Gilli et al models,
Matt et al. 1999).  These calculations are insensitive to the
particular fraction of mid-IR excess galaxies which are actually X-ray
emitters, as the dependence is only upon the stacked X-ray fluxes.

The total background in the 10--30~keV band is estimated to be about 
44~keV~s${}^{-1}$~cm${}^{-2}$~sr$^{-1}$. On the basis of the Gilli et al. (2007) model,
$\approx$~70--80\% of the 10--30 keV background 
can be ascribed to the sources detected
in the deepest X-ray surveys in the
$\approx$~0.5--10~keV band.
Therefore we conclude that this population
of $z=2$ Compton thick AGNs can account for $\sim10-25$\%
of the {\it missing} background in the 10--30~keV band, i.e. of the fraction
due to Compton thick AGNs
which are not individually detected in X-rays.
This is in reasonably good agreement with Gilli
et al.~(2007) model calculations for the background in this energy
range emitted by Compton thick AGNs at $1.4<z<2.5$. The same models
predict that most of the rest of the X-ray background is produced by
similar AGNs at $z<1.4$ (although, Steffen et al. 2007 find marginal
background contributions from X-ray undetected MIPS galaxies 
with \24mu    flux density $>80\mu$Jy at $z\sim1$). 
Our results suggest that the AGNs in these $z<1.4$ galaxies could be
identified from the presence of excess mid-IR emission.

\subsection{On the coeval growth of Black Holes and bulges}

About 20--25\% of the $z\sim 2$ star forming galaxies in GOODS
with $K<22$ (i.e. with stellar masses $\simgt 10^{10}M_\odot$) have
\24mu    excesses by more than 0.5 dex, with the fraction rising to
50--60\% for the most massive star forming galaxies with
$M>4\times10^{10}M_\odot$ (Fig.~\ref{fig:EXC_Mass}).  
We have shown that this appears to be due to the
presence of heavily obscured AGNs in these objects, indicating that
heavily obscured AGN activity was widespread among
star forming galaxies and massive galaxies in general at $z=2$. 
This is different from the local universe, where the fraction of massive
galaxies containing an AGN is only a few percent (Kauffmann et 
al.~2004), supporting the idea that
the BH growth from the most massive BHs must have been
higher in the past (see Fig 2 in Heckman et al. 2004).
The AGN activity is likely connected, in some physical way, to
the intense star formation activity.  In fact, locally a comparably
large fraction of LIRGs and ULIRGs
contain AGNs with detectable mid-IR excess (Genzel et al 1998; Yun et al.~1999
Tran et al 2000; Laurent et al 2000).
Therefore, the
relative fraction of AGNs in LIRGs and ULIRGs
might not be strongly evolving, while the
fraction of AGNs in massive star forming galaxies evolves by a
substantial factor.
Quite obviously, this suggests that the epoch of major build-up of stellar
mass in galaxies at $z\sim 2$ (e.g., Daddi et 
al.~2005b; Papovich et al.~2006; Franceschini et al.~2006) coincided with
a major build-up phase
of the central supermassive BHs, as appear to be the case in the 
submillimeter-emitting galaxy population
(e.g., Alexander et al. 2005a, Borys et al. 2005). This supports the idea
that the local correlations between BH and galaxy
mass were naturally put in place by a concomitant formation.
Is there evidence of such a parallel growth of stellar and BH mass
in our sample of $z\sim 2$ galaxies ?

The rate of BH growth in these objects can be estimated from
their bolometric luminosities, as $L_{\rm bol}=\eta \dot{M}_{\rm
BH}c^2$, with a typical energy conversion efficiency $\eta\approx0.1$
(e.g., Marconi et al.~2004).  Given the large uncertainty in the
bolometric luminosities of the AGNs in mid-IR excess galaxies
($10^{11}\simlt L_{\rm bol}\simlt 10^{12}\;L_\odot$) , we infer that
the average BH accretion rate is in the range $\dot{M}_{\rm
BH}\sim0.05$--0.5$\eta_{0.1}^{-1}\ M_\odot$~yr${}^{-1}$ (here
$\eta_{0.1} = \eta/0.1$).  The mid-IR excess sources are only $\sim
1/4$ of the sample of massive star forming galaxies (i.e., the growth
in stellar mass is happening in a $\sim4$ times larger sample of
galaxies that remain undetected in hard X-rays). The average BH
accretion per galaxy may thus be close to 1/4 of the above estimate.
We are neglecting here the possible additional BH growth inside
galaxies classified in our simplification as mid-IR normal, and BH
growth occurring inside individually X-ray-detected $z=2$ AGNs. The latter
term, from individually hard X-ray detected AGNs inside
$z=2$ massive star forming galaxies,
would increase the BH growth budget by {\em at most} a factor of 2, and likely
much less (see Sect.~4.5).
The massive star forming galaxies in our full GOODS-S sample have average
SFR$\sim70$~M$_\odot$yr${}^{-1}$ (from the UV and soft X-ray). Assuming
explicitly here the more physical IMF of Kroupa (2001) or Chabrier
(2003), and considering (for comparison to the local Universe) that
only some 70\% of these stars will survive at $z=0$, we get a ratio of
$\dot{M}_{\rm BH}/SFR \sim (0.35$--$3.5)\times10^{-3}\eta_{0.1}^{-1}$.

The local BH to galaxy mass correlation implies $M_{\rm
BH}/M_{\rm bulge} \approx 1.5\times10^{-3}$ (McLure \& Dunlop 2001;
Marconi et al.~2004; Ferrarese et al 2006), within about a factor of 2
uncertainty.  Although uncertain, our estimate of the relative growth
rates brackets the observed BH to stellar mass ratio, which
tantalizingly suggests that this population of mid-IR excess galaxies
may indeed represent the major building phase of BHs in massive
galaxies. It is in fact largely expected that massive and star forming
$z=2$ galaxies are the progenitors of local spheroids and bulges
(e.g., Daddi et al 2004ab; Shapley et al 2005; Adelberger et al 2005;
Kong et al. 2006).

From the space density of these sources and their BH accretion rate,
the cosmic BH accretion rate density turns out to be
$10^{-5}-10^{-4}$M$_\odot$yr${}^{-1}$Mpc${}^{-3}$.  This is a factor
of $\sim2.5$--25 larger than the contribution estimated for
$z\approx$~2 bright SMGs (Alexander et~al. 2005a),
which appear to represent the most extreme star-forming galaxies at this
epoch; clearly, fainter submm observations may reveal significantly more 
AGNs and BH growth and eventually overlap with the $K$-selected population
investigated here (which have $S_{850\,\mu m}\sim1$--1.5~mJy, see Daddi
et al 2005b and \hbox{Paper~I}).
However, this accretion-rate density is comparable to the
contribution of luminous quasars at these epochs (e.g.,\ Yu \&
Tremaine 2002;  Alexander et~al. 2005a; Croton et al
2006). Integrating over the 2~Gyr time span corresponding to the
$1.4<z<2.5$ redshift range, this would produce a local BH mass density
of $(0.2$--$2)\times10^{5}$M$_\odot$~Mpc${}^{-3}$.  Comparing to the
estimate of $4.6\times10^5 M_\odot$~Mpc${}^{-3}$ for the local
integrated BH mass density (Marconi et al.~2004), our estimate is
about 5--50\% of the total. This is similar to the estimate that
today's massive galaxies are forming most of their stars in the
$1.4<z<2.5$ redshift range, accounting for $\approx 20$\% of all the
stars in the local Universe (Daddi et al 2005b).

\subsection{AGN duty cycle}

The high incidence of AGNs inside massive
star forming galaxies, 20-30\% for the overall samples, directly implies that
the AGN activity at the luminosity levels at which we detect
it has a relatively long duty cycle, 
at most only 3-4 times shorter that that of star formation. 
If only a fraction of mid-IR excess galaxies are actually X-ray luminous
these estimates would be reduced by the same fraction. Accounting for 
individual X-ray detections of massive star forming galaxies
would instead marginally increase these estimates (Sect.~4.5).

In \hbox{Paper~I} we 
show that massive star forming galaxies with $M>10^{11}M_\odot$ are 
generally ultra luminous IR galaxies (ULIRGs) with a duty cycle of at least
40\% and an estimate time duration of at least 400~Myr. At these high masses
the fraction of mid-IR excess sources reaches $\sim50$\%. Therefore,
our results suggest that this
obscured AGN activity in massive ($M>10^{11}M_\odot$) 
galaxies has a duty cycle of at least
$\sim20$\% and typical durations of at least 200~Myr. 
We emphasize that
these figures are likely to be lower limits because, 
given the downsizing trends,
actively {\em star forming} galaxies are likely to represent a higher fraction
of all massive galaxies at masses lower than $10^{11}M_\odot$ at $z=2$
(i.e., the fraction of quiescent/passive galaxies likely increases 
rapidly with stellar mass at $z=2$). This is consistent with the timescales
inferred by Marconi et al (2004).

It is also worth noting that the strong increase of the fraction of
mid-IR excess galaxies with
stellar mass (see Fig. 5) might imply that the duty cycle of AGN activity
inside star forming galaxies
depends on stellar mass. This could actually be the ultimate
reason for downsizing, i.e. the fact that massive elliptical galaxies
are the first to stop their star formation activity (e.g., Cimatti et
al. 2006; Bundy et al. 2006; Scarlata et al. 2007).  With more massive
systems being more likely to host an AGN, they indeed would be the
first ones to turn passive if AGN activity is the preferred feedback channel
for the  quenching star formation.  Kriek et al (2007) have
proposed a similar argument.

\subsection{Submm galaxies, merging and BH growth}

It is relevant at this point to summarize the comparison between $BzK$ galaxies,
particularly the mid-IR excess objects, and the co-evally selected SMGs, 
as this
has possibly interesting implications for the triggering of BH and galaxy 
growth. We have discussed in Sect.~5.4 that the BH accretion-rate density
at $z=2$ is higher by roughly an order of magnitude
in the mid-IR excess galaxies, over what found in SMGs.
Estimates of the $z=2$ star formation rate density also show that  SMGs
account for a similar share of the overall stellar mass growth.
Compared to the mid-IR excess galaxies selected to $K<22$, 
SMGs are much rarer (at least at bright submm
flux densities of $f_{\rm 850\,\mu m}>5$~mJy; e.g., Coppin et al. 2006), 
have stellar masses larger by roughly a factor 
of 2--3 (e.g., Greve et al 2005),
have star formation rates  about 1 order of magnitude higher
($\approx700$~$M_\odot$~yr$^{-1}$ versus $\approx70$~$M_\odot$~yr$^{-1}$;
Chapman et al. 2005; this paper) but AGN luminosities are quite similar   
(a few $10^{43}$\ergpsec; Alexander et al 2005ab; this paper). This can explain
why generally the SMGs do not display a mid-IR excess 
over their bolometric luminosity inferred from radio (Pope
et al. 2006; see also Menendez-Delmestre et al. 2007;
Valiante et al. 2007), 
as the mid-IR emission from the AGN is likely overshadowed by that
from star formation. This also implies that the ratio of
BH growth to stellar mass growth is higher in mid-IR excess galaxies than 
in SMGs. For the former, the concurrent BH and stellar mass growth
can account for the buildup of galaxies lying on the 
locally defined black-hole--spheroid mass relation,
while this is likely not the case for the activity of SMGs
(Alexander et al. 2005ab).

Submm bright galaxies are likely a {\it special} phase during the buildup
of massive galaxies corresponding to both the highest SFRs
and specific SFRs.
They lie outside the stellar mass versus SFR correlation, as described in 
\hbox{Paper~I}, in the same way that ULIRGs are outliers of the local
stellar mass to SFR correlation. This is consistent with the idea that 
SMGs like local ULIRGs are triggered by major mergers, 
as inferred also on the
basis of their morphologies (Conselice et al 2003; 
Chapman et al 2003), and as suggested also by the compact sizes of the
starbursts observed in CO lines and in the radio (Tacconi 
et al. 2006; Bouche et al 2007). Instead, the fact that $BzK$ galaxies in general define 
a tight stellar-mass to SFR relation argues against an important role
of major mergers in triggering their activity. The long duty cycle 
of their SFR activity (Daddi et al 2005b; \hbox{Paper~I}) also supports this
idea, while SMGs are thought to be comparably much shorted lived
(e.g., Greve et al 2005 but see also Swinbank et al. 2006).

Therefore, an important conclusion that one might derive from our results
is that BH growth in galaxies appear more directly connected with relatively 
stable and long lived star formation, despite still at considerably high rates,
rather than being predominantly 
associated to rare and short lived major mergers events.
We notice that most models of coeval  BH and galaxy growth and of AGN feedback
postulate instead that major mergers are the most relevant events for BH
growths and for triggering feedback (e.g., di Matteo et al. 2005; Schawinski
et al. 2006; Croton et al. 2006). Our result is, in fact, somewhat surprising
given that mergers are known to be very efficient in channeling the
gas toward the center of the galaxies.
Major mergers are likely still required to justify the rapid morphological
transition between massive high-$z$ star forming galaxies with irregular/disk
morphologies into bulge dominated systems (although perhaps disk instabilities
might also play this game; Genzel et al 2006), but it appears that 
a substantial fraction of the BH masses inside local massive galaxies
might have been put in place outside such events. 

\section{AGN feedback via Compton heating ?}
\label{sec:feed}

The widespread presence of obscured AGNs lurking inside massive
star-forming galaxies naturally leads to the question of their
possible feedback effect on their host galaxies.  Feedback from AGNs
is widely believed to be the key factor for solving a major problem of
galaxy formation models based on $\Lambda$CDM cosmology.  In short, these
models require an {\it ad hoc} mechanism to quench star formation in
massive galaxies, which is empirically required to reproduce the
observed color bi-modality of galaxies as well as the old ages of
stars in early-type galaxies. Passively evolving elliptical galaxies
are now known to be present in significant numbers at least up to
$z\sim2.5$ (Cimatti et al.~2004; McCarthy et
al.~2004; Daddi et al.~2005a). As first shown by Granato et al.~(2001;
2004), physically motivated semi-empirical models of galaxy evolution
implementing AGN feedback can successfully account for the existence
of massive and passive galaxies already at early epochs (see also de
Lucia et al.~2006; Springel et al.~2006; Croton et al.~2006; Bower et
al.~2006; Kitzbichler \& White 2006), as shown also in 
\hbox{Paper~I}. Our results are consistent with the 
idea that potentially {\it all} $z=2$ massive star forming galaxies contain 
AGNs, and therefore the {\it ad hoc} mechanism advocated by the models
appears at least to be plausible because the AGNs are indeed there.
Therefore, it would be of great
interest to find a direct observational connection between AGN
activity and star-formation quenching in massive galaxies.

We can imagine two ways that energy can be transferred from the AGN to
the host interstellar medium (ISM), heating it up until it escapes
from the galaxy in a wind. One is via (sub)relativistic jets, and
their possible role in a similar context -- suppression of galaxy {\it
cooling flows} -- has been invoked by, e.g., Binney \& Tabor
(1995). Alternatively (or in addition), the hard X-rays emitted by the
AGN may produce Compton heating as they are absorbed by the ISM, again
powering galactic winds (Ciotti \& Ostriker 1997; 2007). 
AGN jets and accretion related outflows are likely a major channel
for mechanical
feedback in distant galaxies. However, we cannot state anything new on the 
issue based on our sample of $z\sim 2$ galaxies and our results. 
We focus instead
here on the latter option, as the likely Compton thick nature of these
AGNs indicates that Compton heating could indeed be taking place.

The large obscuration of the central BH implies that very high energy
photons are heavily absorbed by the material surrounding the AGN,
which can thus be efficiently heated. The typical star forming galaxy
in our GOODS-S sample has a stellar mass of
$M\sim3\times10^{10}M_\odot$, and the binding energy of the gas can be
estimated to be of order of $M_{gas} \sigma_{\rm v}^2 \approx 3\times 10^{58}
f_{gas}$, where $f_{gas}$ is the gas fraction of the galaxy's baryonic
mass and $\sigma_{\rm v}$ is the velocity dispersion. 
The X-ray luminosity absorbed at high energies, e.g.,
$>2$~keV in the rest-frame, is $\simgt
10^{43}$~erg~s$^{-1}$.  These very high energy photons
can efficiently heat the gas, through Compton scattering.
Assuming, as a limiting case, that all the
absorbed energy at $>2$~keV goes into extracting 
gas from the galactic potential
well, and using $f_{gas}=0.1$, it would take only $\sim
3\times10^6$~yr to blow-off the gas completely. Clearly, the
efficiency is likely to be much lower, chiefly because the material in the BH
surroundings can absorb and re-radiate to lower energies most of the
energy of the absorbed hard X-ray photons. The mid-IR excess itself
testifies for (part of) the AGN energy output being indeed degraded. The
energy available for feedback is the difference between the total hard
X-ray energetic output of the AGN, and the fraction of it which is
re-radiated at lower energies. 
Since both terms are affected by large uncertainties, estimating their
difference would be affected by even larger uncertainties.
Notwithstanding this limitation, the large
unobscured luminosity in hard X-rays that we derive for these objects
argues for the plausibility of a scenario in which Compton heating
plays an important role in quenching star formation.

\section{Future Prospects}
\label{sec:future}

We have discovered a substantial population of heavily obscured,
Compton thick AGNs, residing in a large fraction of massive star
forming galaxies at $z\sim2$.  
We have shown that this result has
important implications for BH--galaxy evolution but our analysis still
leaves open many questions.
There are two crucial points that will need to be
addressed by future studies: a cleaner selection of mid-IR excess
objects and an improved estimate of the obscuring column densities,
hence of $L^{\rm AGN}_{\rm X}$, and of $L^{\rm AGN}_{\rm BOL}$.

For the selection, the main current limitation is that the fiducial
SFR activity of the host galaxy, used to diagnose the presence of an
excess at mid-IR bands, is derived from the UV. Due to the necessity
of large (hence uncertain) reddening corrections, and in some case of UV 
obscuration exceeding what can be inferred from the UV slope, this procedure is
far from being ideal. A better procedure would be to derive a reliable
measurement of the bolometric IR luminosity of the galaxy from 
long wavelength observations, and
compare this to the mid-IR rest-frame luminosity. The availability of
ALMA, SCUBA2 and especially Herschel observations in the near future
should all substantially help with this task, providing much better
multiwavelength SEDs for the dust emission, from mid-IR wavelengths
through the far-IR peak and beyond. With such a detailed SED, one
could try to disentangle the cold-dust and the warm dust emission, due
to, respectively, star formation and AGN heating.

Measuring reliable values for $N_{\rm H}$, determining accurately what
fraction of mid-IR excess sources are Compton thick, and perhaps also
detecting individually these sources, would require extremely
sensitive observations. With {\it Chandra} this could require an
ultra-deep integration, of order of \hbox{5-10~Ms}, to allow to detect some of
the brightest objects. Perspectives also look promising
for the next generation of
X-ray satellites currently under discussion, such as XEUS (Bavdaz et al 2006);
if realized with a fairly high angular resolution ($\sim2$--3$''$) 
XEUS would be ideally suited to X-ray spectroscopy of these X-ray faint AGNs. 
Rest-frame optical spectroscopy can also provide
insights by looking for luminous high-excitation emission lines such as
[OIII]$\lambda$5007.

Mid-IR spectroscopy will also be crucial to improve our understanding of these
sources.
We have not as yet explicitly demonstrated that AGN emission
does account for the entirety of the mid-IR excess flux density.
Our analysis cannot exclude the possibility that 
intrinsically different mid-IR SED properties, e.g.~anomalously strong
PAHs, exist
in a fraction of $z=2$ galaxies, with the mid-IR still powered
by SFR. This might be related to the presence of different SFR modes, 
rare or non-existent by $z<1$, a result that would be quite interesting
on its own.
Clearly, the ideal means to obtain a direct diagnostic 
of the cause of the \24mu    excess would be to perform
spectroscopy at these wavelengths with {\it Spitzer} and {\it Akari}.
Given that the typical mid-IR excess
galaxy is highly star forming, we do expect to see PAH features in a large
majority of the mid-IR excess galaxies. However, 
if the mid-IR excess is directly 
produced by the AGN, we should detect also an important contribution from AGN
continuum in most of the mid-IR spectra, lowering the equivalent width 
of the PAH emission features.
As mid-IR excess objects even at $z\sim2$ are relatively
bright at \24mu, with flux densities often in excess of 100--200$\mu$Jy, 
the observations are feasible and useful spectra 
are already available in the {\it Spitzer} archive, 
although their analysis exceeds the scope of the present paper.

\section{Summary and conclusions}
\label{sec:end}

We have studied the nature of a population of $z\sim2$ galaxies that
displays a marked excess flux density at mid-IR wavelengths, over what
expected to arise from just normal star formation activity as
estimated from their rest-frame UV. The examined sample includes over
600 $K$-selected $z\sim2$ galaxies in the two GOODS fields 
(with 25\% of them having a
spectroscopic redshifts), with multiwavelength informations including
data in UV, optical, near-IR, Spitzer~IRAC (3.6--8.0$\mu$m bands), Spitzer
MIPS (\24mu    and 70\,$\mu$m bands), SCUBA, VLA 1.4~Ghz, and X-ray from
{\it Chandra}. This sample has been cleaned by all known
AGNs, identified via their {\it Chandra} 2--8~keV X-ray
emission and/or by the presence of a power-law SED over the IRAC
bands.  The major results of this work can be summarized as follow.

\begin{itemize}

\item Roughly 20--30\% of the massive star forming galaxies at $z\sim2$,
selected down to $K<22$, display a mid-IR excess larger than
about a factor of 3, which we classify as ``mid-IR excess
galaxies". This fraction increases rapidly with the stellar mass of
the galaxies, reaching $\sim50$--60\% for $M>4\times10^{10}M_\odot$.  We
have shown that our results do not strongly depend on the exact
threshold used for defining these mid-IR excess galaxies.

\item This population of mid-IR excess galaxies is at slightly higher
average redshift than normal galaxies ($<\! z\! >=2.1$ vs. 1.9), but
overall the two redshift distributions are similar.  This population
becomes dominant at the highest 8\,$\mu$m (rest-frame) luminosities
(for $L(8\mu m)>10^{11}L_\odot$), that we accurately derive at $z\sim 2$
from Spitzer~MIPS \24mu  flux densities. Relatively shallow survey with 
Spitzer MIPS of the $z\sim2$ Universe are thus the ones that are proportionally
most affected by the presence of mid-IR excess galaxies.

\item Mid-IR excess sources have redder $K-5.8$ colors than normal
galaxies, reflecting different intrinsic SEDs in the optical to
near-IR rest-frame, that can not be attributed to mere redshift
effects. For these mid-IR excess sources, the peak of the stellar
dominated part of the SED is longward of the canonical 1.6$\mu$m rest
frame. We interpret this as evidence for the presence of a warm dust AGN
contribution to the continuum in these sources, consistent also with the
finding that the excess flux density seems to become weaker beyond
20$\mu$m rest-frame, based on Spitzer 70\,$\mu$m observations.

\item This excess AGN component over the near-IR rest-frame will likely
bias the estimates of stellar masses and photometric redshifts for
some of the $z\sim2$ galaxies, when including IRAC photometry in the
SEDs.  However, we quantify the effect on the derived photometric redshifts 
to be only at the level of 2-3\% in $(1+z)$, for the mid-IR excess
sources.

\item Stacking of {\it Chandra} X-ray data reveals a strong dichotomy
in the two populations: the normal sources have a soft spectrum,
consistent with star formation in both 0.5--2 and 2--8~keV bands. 
Contrary to them,
mid-IR excess objects exhibit a much harder spectrum, unambiguously
showing the presence of highly obscured AGNs inside these sources.

\item We use a variety of methods to infer $N_{\rm H}$, the column
density of the absorbing material along the line of sight. Our best
estimate suggests typical obscuration of $N_{\rm H} = 10^{24.5}$
cm$^{-2}$, derived from the X-ray spectrum
after accounting for the X-ray emission due to star formation. 
As a result, we infer that the majority of the
mid-IR excess galaxies is likely to host a Compton thick AGN.

\item This previously unknown population of AGNs has typical X-ray
luminosity of $(1$--$4)\times10^{43}$~erg~s$^{-1}$ in the 2--8~keV rest
frame band. We infer a range for the bolometric
luminosity of $10^{11}-10^{12}L_\odot$. This is
comparable to the bolometric luminosities of their host galaxies, as due
to star formation, for which we estimate a typical value of
$\sim5\times10^{11}L_\odot$.

\item The sky density of Compton thick AGNs inside massive star
 forming galaxies to $K<22$ is $\sim 3200$~deg$^{-2}$ and their space
 density is $\sim2.6\times10^{-4}$~Mpc$^{-3}$.  Their space density is
 higher than that of all AGNs already known at $z=2$, and much higher
 than that of previously known Compton thick populations at high
 redshifts.  The space density of Compton thick AGNs that we derive
 agrees reasonably well with that predicted by the background
 synthesis models of Gilli et al (2007).

\item Despite their large space density, we extrapolate that this
population can account for only about 10--25\% of the {\em missing} X-ray
background in the observed 10--30~keV band, in
agreement again with the prediction of Gilli et al (2007). Most of the
high energy background still to be resolved is likely inside similar
sources but residing at $z\simlt 1.4$ rather than at $z\sim2$.

\item The widespread presence of obscured AGNs inside massive star
forming galaxies at $z\sim 2$ seems to quite naturally account for the
concurrent growth of the BH together with the stellar mass of
the hosting galaxies. Within quite large quantitative uncertainties,
this picture is consistent with these objects approaching the local
values for the ratios between the mass of central BH and that
of their hosting bulges.

\item A relatively long duty cycle for the AGN activity at $z=2$, 
only a factor of 3-4 lower than that of star formation activity, is suggested 
by the widespread presence of AGNs that appear to be present
among massive star forming galaxies.

\item Comparing to submm selected galaxies (SMGs), we find that
BHs are growing faster in mid-IR excess galaxies, relatively to the ongoing
SFR, and their integrated contribution to the BH accretion-rate density
are also larger. This suggest that of BH growth in massive galaxies is taking
place outside major merger events.\\

\item The large, increasing incidence of Compton thick AGNs with
stellar mass, along with the energy deposition provided by the
absorption of the hard X-ray photons, suggest that these objects may
be responsible for the AGN feedback eventually leading to the
termination of the star formation activity, and to do so beginning
with the most massive galaxies as in the {\it downsizing}
scenario. Although existing data do not allow to firmly prove this
tantalizing possibility, we show that the involved energetics is
consistent with this picture.

\item We finally discuss future prospects for improving our still
limited understanding of the properties of mid-IR excess galaxies and
Compton thick AGNs at high redshift. The crucial and obvious required
next step is the full construction of their spectral energy
distribution, including ALMA, SCUBA-2, and Herschel facilities as they
become available, as well as of their {\em spectral} mid-IR properties, via
Spitzer IRS and Akari spectroscopy. Extremely deep {\it
Chandra} (5-10 Ms) exposures would be required to directly
detect some of these Compton thick AGNs.

\end{itemize}

In summary, we have uncovered a major population of luminous and obscured
$z=2$ AGNs, previously largely unknown.
With AGN and star formation activity in massive galaxies
peaking at $z\sim 2$, this newly detected population of Compton thick
AGNs appears to be an excellent candidate for playing a major role
both for establishing the BH mass--bulge mass relation, and for
providing the feedback necessary to discontinue further star formation
in massive galaxies.

\acknowledgements
We thank the rest of the GMASS team for allowing us to use
the still unpublished spectroscopic redshifts, and the the many 
other members of the GOODS team,
who have helped to make these observations possible.
We are grateful to Emily MacDonald, Daniel Stern and Hy Spinrad
for collecting some of the redshifts
used in this work. 
We thank Gianni Zamorani and Susanne Madden for useful
comments and discussions.
ED gratefully acknowledges NASA support (at the beginning of this work)
through the Spitzer
Fellowship Program, award 1268429.
DMA thanks the Royal Society for funding. RG acknowledges 
financial support from the Italian Space Agency (ASI) under the contract 
ASI-INAF I/023/05/0. 
WNB acknowledges Spitzer Space Telescope grant 1278940.
JK acknowledges financial support from the German Science Foundation (DFG)
under contract SFB-439.
Support for this work, part of the Spitzer Space Telescope 
Legacy Science Program, was provided by NASA, Contract Number 1224666
issued by the JPL, Caltech, under NASA contract 1407.


\begin{thebibliography}{}
\bibitem[~]{2005ApJ...620L..75A} Adelberger K.~L., Erb D.~K., Steidel C.~C., Reddy N.~A., Pettini M., Shapley A.~E., 2005, ApJ, 620, L75 
\bibitem[~]{2003AJ....126..539A} Alexander, D.~M., et al.\ 2003a, \aj, 126, 539
\bibitem[~]{2003AJ....125..383A} Alexander D.~M., et al., 2003b, AJ, 125, 383 
\bibitem[~]{2005Natur.434..738A} Alexander D.~M., Smail I., Bauer F.~E., Chapman S.~C., Blain A.~W., Brandt W.~N., Ivison R.~J., 2005a, Natur, 434, 738 
\bibitem[~]{2005ApJ...632..736A} Alexander D.~M., Bauer F.~E., Chapman S.~C., Smail I., Blain A.~W., Brandt W.~N., Ivison R.~J., 2005b, ApJ, 632, 736 
\bibitem[~]{2006ApJ...640..167A} Alonso-Herrero A., et al., 2006, ApJ, 640, 167 
\bibitem[~]{1993ARA&A..31..473A} Antonucci R., 1993, ARA\&A, 31, 473
\bibitem[~]{2007ApJ...656..148A} Armus L., et al., 2007, ApJ, 656, 148 
\bibitem[~]{avni}Avni Y., 1976, ApJ, 210, 642
\bibitem[~]{2007ApJ...654..764B} Barger A.~J., Cowie L.~L., Wang W.-H., 2007, ApJ, 654, 764 
\bibitem[~]{2006SPIE.6266E..51B} Bavdaz M., et al., 2006, SPIE, 6266, 51
\bibitem[~]{Binn} Binney, J., \& Tabor, G. 1995, MNRA, 276, 663
\bibitem[~]{2005ApJ...635..853B} Borys C., Smail I., Chapman S.~C., Blain A.~W., Alexander D.~M., Ivison R.~J., 2005, ApJ, 635, 853 
\bibitem[~]{bouche} Bouche N., Cresci G., Davies R., et al., 2007, ApJ in press (arXiv:0706.2656)
\bibitem[~]{Bower} Bower, R.G., et al. 2006, MNRAS, 370, 654
\bibitem[~]{2005ARA&A..43..827B} Brandt W.~N., Hasinger G., 2005, ARA\&A, 43, 827 
\bibitem[~]{2005A&A...432...69B} Brusa M., et al., 2005, A\&A, 432, 69 
\bibitem[~]{Bund} Bundy, K, et al. 2006, ApJ, 651, 120
\bibitem[~]{2000ApJ...533..682C} Calzetti D., Armus L., Bohlin R.~C., et al., 2000, ApJ,  533, 682
\bibitem[~]{nat} Cimatti A., Daddi E., Renzini A., et al., 2004, Nature, 430, 184
\bibitem[~]{2006A&A...453L..29C} Cimatti A., Daddi E., Renzini A., 2006, A\&A, 453, L29 
\bibitem[~]{Ciot1} Ciotti, L., \& Ostriker, J.P. 1997,  ApJ, 487, L10
\bibitem[~]{Ciot2} Ciotti, L., \& Ostriker, J.P. 2007,  submitted to ApJ (astro-ph/0703057)
\bibitem[~]{2003ApJ...599...92C} Chapman S.~C., Windhorst R., Odewahn S., Yan H., Conselice C., 2003, ApJ, 599, 92 
\bibitem[~]{2005ApJ...622..772C} Chapman, S.~C., Blain, A.~W., Smail, I., \& Ivison, R.~J.\ 2005, \apj, 622, 772
\bibitem[~]{cha07} Chartas G., Brandt W.~N., Gallagher S.~C., Proga D., 2007, AJ, 133, 1849 
\bibitem[~]{1995A&A...296....1C} Comastri A., Setti G., Zamorani G., Hasinger G., 1995, A\&A, 296, 1 
\bibitem[~]{2003ApJ...596L...5C} Conselice C.~J., Chapman S.~C., Windhorst R.~A., 2003, ApJ, 596, L5 
\bibitem[~]{2006MNRAS.372.1621C} Coppin K., et al., 2006, MNRAS, 372, 1621 
\bibitem[~]{2004MNRAS.349.1397C} Croom S.~M., Smith R.~J., Boyle B.~J., Shanks T., Miller L., Outram P.~J., Loaring N.~S., 2004, MNRAS, 349, 1397 
\bibitem[~]{cro} Croton D.~J., et al., 2006, MNRAS, 365, 11
\bibitem[~]{2004ApJ...600L.127D} Daddi E., et al., 2004a, ApJ, 600, L127 
\bibitem[~]{2004ApJ...617..746D} Daddi E., Cimatti A., Renzini A., Fontana A., Mignoli M., Pozzetti L., Tozzi P., Zamorani G., 2004b, ApJ, 617, 746
\bibitem[~]{2005ApJ...626..680D} Daddi E., et al., 2005a, ApJ, 626, 680 
\bibitem[~]{d05b} Daddi E., Dickinson M., Chary R., et al, 2005b, ApJ, 631, L13
\bibitem[~]{PI} Daddi E., Dickinson M., Morrison G., et al., 2007, ApJ in press (arXiv:0705.2831) (\hbox{Paper~I})
\bibitem[~]{2006ApJ...646..161D} Dale D.~A., et al., 2006, ApJ, 646, 161 
\bibitem[~]{2006MNRAS.366..499D} De Lucia G., Springel V., White S.~D.~M., Croton D., Kauffmann G., 2006, MNRAS, 366, 499
\bibitem[~]{2005Natur.433..604D} Di Matteo T., Springel V., Hernquist L., 2005, Natur, 433, 604 
\bibitem[~]{2007ApJ...660..167D} Donley J.~L., Rieke G.~H., P{\'e}rez-Gonz{\'a}lez P.~G., Rigby J.~R., Alonso-Herrero A., 2007, ApJ, 660, 167
\bibitem[~]{elb07} Elbaz D., Daddi E., Le Borgne D., et al., 2007, A\&A in press (astro-ph/0703653)
\bibitem[~]{1994ApJS...95....1E} Elvis M., et al., 1994, ApJS, 95, 1 
\bibitem[~]{2000ApJ...539L...9F} Ferrarese L., Merritt D., 2000, ApJ, 539, L9 
\bibitem[~]{Ferr} Ferrarese, L., et al. 2006, ApJ, 644, L21
\bibitem[~]{Fiore} Fiore F., et al., 2007, ApJ, in press (arXiv:0705.2864)
\bibitem[~]{2006A&A...453..397F} Franceschini A., et al., 2006, A\&A, 453, 397 
\bibitem[~]{fray06} Frayer D.T., Huynh M.T., Chary R., et al., 2006, ApJ, 647, L9
\bibitem[~]{2000ApJ...539L..13G} Gebhardt K., et al., 2000, ApJ, 539, L13 
\bibitem[~]{1998ApJ...498..579G} Genzel R., et al., 1998, ApJ, 498, 579 
\bibitem[~]{2006Natur.442..786G} Genzel R., et al., 2006, Natur, 442, 786 
\bibitem[~]{2002ApJS..139..369G} Giacconi R., Zirm A., Wang J., et al., 2002, ApJS,  139, 369
\bibitem[~]{giava} Giavalisco M., Ferguson H.C., Koekemoer A., et al., 2004, ApJ. 600, L93
\bibitem[~]{2001A&A...366..407G} Gilli R., Salvati M., Hasinger G., 2001, A\&A, 366, 407 
\bibitem[~]{2007A&A...463...79G} Gilli R., Comastri A., Hasinger G., 2007, A\&A, 463, 79 
\bibitem[~]{grana} Granato, G.L. 2001, MNRAS, 324, 527
\bibitem[~]{2004ApJ...600..580G} Granato, G.~L., De Zotti, G., Silva, L., Bressan, A., \& Danese, L.\ 2004, \apj, 600, 580
\bibitem[~]{2005MNRAS.359.1165G} Greve T.~R., et al., 2005, MNRAS, 359, 1165 
\bibitem[~]{2001A&A...365L..45H} Hasinger G. 2004. Nucl. Phys. B (Proc. Suppl.) 132, 86
\bibitem[~]{2005A&A...441..417H} Hasinger G., Miyaji T., Schmidt M., 2005, A\&A, 441, 417 
\bibitem[~]{xx-max} Hickox R.C. \& Markevitch M., 2007, ApJL in press (astro-ph/0702556)
\bibitem[~]{2007ApJ...654..731H} Hopkins P.~F., Richards G.~T., Hernquist L., 2007, ApJ, 654, 731 
\bibitem[~]{2005AJ....129...86H} Hornschemeier A.~E., Heckman T.~M., Ptak A.~F., Tremonti C.~A., Colbert E.~J.~M., 2005, AJ, 129, 86 
\bibitem[~]{Kitz} Kitzbichler, M.G. \& White, S.D.M. 2006, MNRAS, 366, 858
\bibitem[~]{kong} Kong X., Daddi E., Arimoto N., et al., 2006, ApJ, 638, 72
\bibitem[~]{1995ARA&A..33..581K} Kormendy J., Richstone D., 1995, ARA\&A, 33, 581 
\bibitem[~]{2kriek} Kriek M., van Dokkum P.G., Franx M., et al., 2007, submitted to ApJ (astro-ph/0611724)
\bibitem[~]{2005ApJ...625...89K} Krivonos R., Vikhlinin A., Churazov E., Lutovinov A., Molkov S., Sunyaev R., 2005, ApJ, 625, 89 
\bibitem[~]{2004ApJS..154..166L} Lacy M., et al., 2004, ApJS, 154, 166 
\bibitem[~]{2005ApJ...635..864L} La Franca F., et al., 2005, ApJ, 635, 864 
\bibitem[~]{A&A...359..887L} Laurent O., Mirabel I.~F., Charmandaris V., Gallais P., Madden S.~C., Sauvage M., Vigroux L., Cesarsky C., 2000, A\&A, 359, 887 
\bibitem[~]{2004A&A...418..465L} Lutz D., Maiolino R., Spoon H.~W.~W., Moorwood A.~F.~M., 2004, A\&A, 418, 465
\bibitem[~]{1998AJ....115.2285M} Magorrian J., et al., 1998, AJ, 115, 2285 
\bibitem[~]{2004MNRAS.351..169M} Marconi A., Risaliti G., Gilli R., Hunt L.~K., Maiolino R., Salvati M., 2004, MNRAS, 351, 169 
\bibitem[~]{1980ApJ...235....4M} Marshall F.~E., Boldt E.~A., Holt S.~S., Miller R.~B., Mushotzky R.~F., Rose L.~A., Rothschild R.~E., Serlemitsos P.~J., 1980, ApJ, 235, 4 
\bibitem[~]{2006MNRAS.370.1479M} Mart{\'{\i}}nez-Sansigre A., Rawlings S., Lacy M., Fadda D., Jarvis M.~J., Marleau F.~R., Simpson C., Willott C.~J., 2006, MNRAS, 370, 1479 
\bibitem[~]{Matt} Matt et al. 1999, A\&A, 341, L39
\bibitem[~]{2004ApJ...614L...9M} McCarthy P.~J., et al., 2004, ApJ, 614, L9
\bibitem[~]{2001MNRAS.327..199M} McLure R.~J., Dunlop J.~S., 2001, MNRAS, 327, 199 
\bibitem[~]{2007ApJ...655L..65M} Men{\'e}ndez-Delmestre K., et al., 2007, ApJ, 655, L65 
\bibitem[~]{noe1} Noeske K.G., Weiner B.J., Faber S.M., et al., 2007a, ApJ Letters, in press (astro-ph/0701924)
\bibitem[~]{DRGs} Papovich C., et al. 2006, ApJ, 640, 92
\bibitem[~]{P07} Papovich C., Rudnick G., Le Floc'h E., et al., 2007, ApJ, in press (arXiv:0706.2164)
\bibitem[~]{galfit} Peng C.~Y., Ho L.~C., Impey C.~D., Rix H.-W., 2002, AJ, 124, 266
\bibitem[~]{2002A&A...382..843P} Persic M., Rephaeli Y., 2002, A\&A, 382, 843 
\bibitem[~]{2006ApJ...642..673P} Polletta M.~d.~C., et al., 2006, ApJ, 642, 673 
\bibitem[~]{2005MNRAS.358..149P} Pope A., Borys C., Scott D., Conselice C., Dickinson M., Mobasher B., 2005, MNRAS, 358, 149
\bibitem[~]{2006MNRAS.370.1185P} Pope A., et al., 2006, MNRAS, 370, 1185 
\bibitem[~]{Pozzi} Pozzi F., Vignali C., Comastri A., et al., 2007, submitted to ApJ (arXiv:0704.0735) 
\bibitem[~]{2003A&A...399...39R} Ranalli P., Comastri A., Setti G., 2003, A\&A,  399, 39
\bibitem[~]{2004ApJS..154..160R} Rigby J.~R., et al., 2004, ApJS, 154, 160
\bibitem[~]{2002AJ....124.3050S} Sawicki M., 2002, AJ, 124, 3050 
\bibitem[~]{Scar} Scarlata, C., et al. 2007, ApJS (in press, astro-ph/0701746
\bibitem[~]{2006Natur.442..888S} Schawinski K., et al., 2006, Natur, 442, 888 
\bibitem[~]{1968ApJ...151..393S} Schmidt M., 1968, ApJ, 151, 393 
\bibitem[~]{shapley} Shapley A., et al., 2005, ApJ, 626, 698
\bibitem[~]{1998A&A...331L...1S} Silk J., Rees M.~J., 1998, A\&A, 331, L1 
\bibitem[~]{2004MNRAS.355..973S} Silva L., Maiolino R., Granato G.~L., 2004, MNRAS, 355, 973 
\bibitem[~]{sperg03} Spergel D.N., et al., 2007, ApJ, in press (astro-ph/0603449)
\bibitem[~]{spri.620L..79S} Springel V., Di Matteo T., Hernquist L., 2005, ApJ, 620, L79 
\bibitem[~]{2006Natur.440.1137S} Springel V., Frenk C.~S., White S.~D.~M., 2006, Natur, 440, 1137 
\bibitem[~]{Steff} Steffen A.T., Brandt W.N., Alexander D.M., Gallagher S.C., Lehmer B.D., 2007, submitted to ApJ (arXiv:0705.2213)
\bibitem[~]{2005ApJ...631..163S} Stern D., et al., 2005, ApJ, 631, 163 
\bibitem[~]{2006MNRAS.371..465S} Swinbank A.~M., Chapman S.~C., Smail I., Lindner C., Borys C., Blain A.~W., Ivison R.~J., Lewis G.~F., 2006, MNRAS, 371, 465 
\bibitem[~]{2006A&A...451..457T} Tozzi P., et al., 2006, A\&A, 451, 457 
\bibitem[~]{2001ApJ...552..527T} Tran Q.~D., et al., 2001, ApJ, 552, 527
\bibitem[~]{2004ApJ...616..123T} Treister E., et al., 2004, ApJ, 616, 123 
\bibitem[~]{2003ApJ...598..886U} Ueda Y., Akiyama M., Ohta K., Miyaji T., 2003, ApJ, 598, 886 
\bibitem[~]{1995PASP..107..803U} Urry, C.~M., \& Padovani, P., 1995, PASP, 107, 803 
\bibitem[~]{2007ApJ...660.1060V} Valiante E., Lutz D., Sturm E., Genzel R., Tacconi L.~J., Lehnert M.~D., Baker A.~J., 2007, ApJ, 660, 1060 
\bibitem[~]{1999A&A...349L..57V} Vignati P., et al., 1999, A\&A, 349, L57 
\bibitem[~]{2005MNRAS.357.1281W} Worsley M.~A., et al., 2005, MNRAS, 357, 1281 
\bibitem[~]{2006MNRAS.368.1735W} Worsley M.~A., Fabian A.~C., Bauer F.~E., Alexander D.~M., Brandt W.~N., Lehmer B.~D., 2006, MNRAS, 368, 1735 
\bibitem[~]{1999Ap&SS.266...29Y} Yun M.~S., Hibbard J.~E., Condon J.~J., Reddy N., 1999, Ap\&SS, 266, 29 
\bibitem[~]{2004ApJS..155...73Z} Zheng W., et al., 2004, ApJS, 155, 73 
\end{thebibliography}
\end{document}